\documentclass[manuscript,nonacm,screen]{acmart}

\usepackage{amsmath,amsthm,amsbsy}
\usepackage{microtype}
\usepackage{algorithm}
\usepackage{setspace}
\usepackage{color}
\usepackage{xcolor}
\usepackage[noend]{algpseudocode}
\usepackage{multirow}
\usepackage{paralist}
\usepackage{xspace}

\let\originalleft\left
\let\originalright\right
\renewcommand{\left}{\mathopen{}\mathclose\bgroup\originalleft}
\renewcommand{\right}{\aftergroup\egroup\originalright}

\newcommand{\ra}{\rightarrow}

\renewcommand{\Pr}{\operatorname*{\textbf{\textup{Pr}}}}
\DeclareMathOperator*{\II}{\textbf{\textup{I}}}

\DeclareMathOperator*{\HH}{\textbf{\textup{H}}}

\DeclareMathOperator*{\EE}{\textbf{\textup{E}}}

\renewcommand{\ge}{\geqslant}

\renewcommand{\le}{\leqslant}
\newcommand{\set}[1]{\{ #1 \}}

\newcommand{\lb}{\left}
\newcommand{\rb}{\right}
\newcommand{\lt}{\left}
\newcommand{\rt}{\right}

\DeclareMathOperator{\polylog}{\ensuremath{\mathrm{polylog}}}
\DeclareMathOperator{\poly}{\ensuremath{\mathrm{poly}}}

\newcommand{\congest}{\ensuremath{\mathsf{CONGEST}}}

\makeatletter
\newtheorem*{rep@theorem}{\rep@title}
\newcommand{\newreptheorem}[2]{%
\newenvironment{rep#1}[1]{%
\def\rep@title{#2 \ref{##1}}%
\begin{rep@theorem}[restated]}%
{\end{rep@theorem}}}
\makeatother
\newreptheorem{lemma}{Lemma}
\newreptheorem{theorem}{Theorem}

\newboolean{short}
\setboolean{short}{false}
\newcommand{\onlyShort}[1]{\ifthenelse{\boolean{short}}{#1}{}}
\newcommand{\onlyLong}[1]{\ifthenelse{\boolean{short}}{}{#1}}

\theoremstyle{definition}

\newtheorem{definition}{Definition}

\theoremstyle{plain}
\newtheorem{lemma}{Lemma}
\newtheorem{theorem}{Theorem}

\newtheorem{corollary}{Corollary}
\theoremstyle{plain}

\newtheorem{open_problem}{Open Problem}

\newcommand{\cA}{\mathcal{A}}

\newcommand{\cD}{\mathcal{D}}

\newcommand{\cG}{\mathcal{G}}
\newcommand{\cR}{\mathcal{R}}
\newcommand{\cS}{\mathcal{S}}

\newcommand{\md}{\middle}
\newcommand{\ann}[1]{%
\text{\footnotesize(#1)}\quad}

\usepackage{mathtools}

\usepackage{tikz}
\usetikzlibrary{
  graphs,
  graphs.standard
}

\setboolean{short}{true}

\newcommand{\local}{\ensuremath{\mathsf{LOCAL}}}

\newcommand{\ktone}{\ensuremath{\mathsf{KT_1}}}
\newcommand{\ktzero}{\ensuremath{\mathsf{KT_0}}}

\newcommand{\critical}{{C}}
\DeclareMathOperator{\lic}{\mathsf{LIC}}
\newcommand{\dense}{\mathsf{Dense}}

\newcommand{\sparse}{S}

\DeclareMathOperator{\CC}{\mathsf{CC}}

\usepackage{cases}
\usepackage[most]{tcolorbox}
\tcbuselibrary{theorems}
\newtcbtheorem[number within=section]{Definition}{Definition}{
  enhanced,
  sharp corners,
  attach boxed title to top left={
    yshifttext=-1mm
  },
  colback=white,
  colframe=gray!75!white,
  fonttitle=\bfseries,
  boxed title style={
    sharp corners,
    size=small,
    colback=gray!75!white,
    colframe=gray!75!white,
  }
}{def}

\ccsdesc[500]{Theory of computation~Models of computation}
\ccsdesc[500]{Theory of computation}
\ccsdesc[500]{Theory of computation~Distributed algorithms}

\acmBooktitle{}
\title[]{The Local Information Cost of Distributed Graph Spanners}

 \author{Peter Robinson}
 \authornote{A short version of this paper appeared in the proceedings of the 32th ACM-SIAM Symposium on Discrete Algorithms (SODA'21).
}
\affiliation{
  \department{School of Computer \& Cyber Sciences}
  \institution{Augusta University}
  \country{USA}
}

\begin{document}
\begin{abstract}
  We introduce the \emph{local information cost} ($\lic$), which quantifies the amount of information that nodes in a network need to learn when solving a graph problem.
We show that the local information cost presents a natural lower bound on the communication complexity of distributed algorithms.
For the synchronous $\congest$ $\ktone$ model, where each node has initial knowledge of its neighbors' IDs, we prove that $\Omega\lb(\frac{\lic_\gamma(P)}{\log\tau \log n}\rb)$ bits are required for solving a graph problem $P$ with a $\tau$-round algorithm that errs with probability at most $\gamma$.
Our result is the first lower bound that yields a general trade-off between communication and time for graph problems in the $\congest$ $\ktone$ model.

We demonstrate how to apply the local information cost by deriving a lower bound on the communication complexity of computing a spanner with multiplicative stretch $2t-1$ that consists of at most $O(n^{1+\frac{1}{t} + \epsilon})$ edges, where $\epsilon = O\lb( {1}/{t^2} \rb)$.
More concretely, we show that any $O(\poly(n))$-time spanner algorithm must send at least $\tilde\Omega\lb(\tfrac{1}{t^2} n^{1+{1}/{2t}}\rb)$ bits.
Previously, only a trivial lower bound of $\tilde \Omega(n)$ bits was known for this problem. 

A consequence of our lower bound is that achieving both time- and communication-optimality is impossible when designing a distributed spanner algorithm.
In light of the work of King, Kutten, and Thorup (2015), this shows that computing a minimum spanning tree can be done significantly faster than finding a spanner when considering algorithms with $\tilde O(n)$ communication complexity.
Our results also imply time complexity lower bounds for these problems in the node-capacitated clique of Augustine, Ghaffari, Gmyr, Hinnenthal, Scheideler, Kuhn, and Li (2019), and in the push-pull gossip model with limited bandwidth, as studied by Haeupler, Mohapatra, and Su (2018).

\end{abstract}
\maketitle

\section{Introduction}
\label{sec:intro}

Designing fast and communication-efficient distributed algorithms is crucial for many applications.
Modern-day examples include building large-scale networks of resource-restricted devices and processing massive data sets in a distributed system.
When analyzing the performance of distributed algorithms, communication efficiency is usually quantified by the message complexity, i.e., the total number of messages sent by the algorithm, and the communication complexity, which refers to the total number of bits sent throughout the execution.
Recently, there has been significant interest in obtaining communication-efficient algorithms for solving fundamental graph problems in the message passing setting.
Due to \cite{jacm15}, it is known that, even for very basic problems such as single-source broadcast and constructing a spanning tree, $\Omega(m)$ messages (and bits) are required in an $n$-node graph $G$ that has $m$ edges, assuming that nodes are initially unaware of the IDs of their neighbors and messages are addressed using port numbers rather than specific IDs.
This model is called the \emph{clean network model}~\cite{peleg_book} or \emph{port numbering model}~\cite{jukkaLocalAlgorithms} (if nodes do not have IDs), and several time- and communication-optimal algorithms have been obtained that match the $\Omega(m)$ barrier, e.g., for minimum spanning trees (MST) \cite{DBLP:conf/stoc/Pandurangan0S17,DBLP:conf/podc/Elkin17}, approximate single-source shortest paths~\cite{DBLP:conf/podc/HaeuplerHW18}, and leader election~\cite{jacm15}.

Since the clean network model does not capture the more realistic setting of IP networks where nodes are aware of the IDs of their peers initially, there has been a growing interest in studying communication-efficient algorithms under the \emph{$\ktone$ assumption}~\cite{DBLP:journals/jacm/AwerbuchGPV90,peleg_book}, where nodes have unique IDs of length $\Theta(\log n)$ and each node knows the IDs of its neighbors in $G$ from the start.
\cite{DBLP:conf/podc/KingKT15} were the first to present a minimum spanning tree (MST) algorithm that runs in $\tilde O(n)$ rounds and sends $\tilde O(n)$ bits in the $\congest$ model~\cite{peleg_book} under the $\ktone$ assumption by using linear graph sketching techniques.
More recently, new distributed algorithms have been proposed for several graph problems in this setting; for instance, algorithms for single-source broadcast~\cite{DBLP:conf/wdag/GhaffariK18a}, graph verification problems~\cite{DBLP:conf/wdag/GmyrP18}, and computing an MST in the asynchronous setting~\cite{DBLP:conf/icdcn/MashreghiK17,DBLP:conf/wdag/MashreghiK19} have been obtained that achieve a message complexity of $o(m)$.
A second motivation for studying communication complexity stems from the area of distributed large-scale graph processing:
For instance, in \cite{soda15} it is shown how to convert any time- \emph{and} communication-efficient $\congest$ $\ktone$ algorithm to a fast $k$-machine algorithm.

A somewhat counterintuitive feature of the $\ktone$ assumption is that it allows solving \emph{any} graph problem using only $\tilde O(n)$ bits of communication by leveraging silence to convey information, albeit at the cost of increasing the time complexity to an exponential number of rounds:
If we first construct a rooted spanning tree using the algorithm of \cite{DBLP:conf/podc/KingKT15}, we can then employ a simple time-encoding scheme using $\tilde O(n)$ bits to collect the entire graph topology at the root, who locally solves the problem. 
Subsequently, the root can disseminate the result of the computation (no matter the output size) to all nodes again via time-encoding.\footnote{For completeness, we provide a more detailed description of this time-encoding procedure in Appendix~\ref{sec:time-encoding}.}
Consequently, any lower bound on the required communication must be conditional on the algorithm terminating in a sufficiently small number of rounds; note that this stands in contrast to the clean network model where the $\Omega(m)$ barrier holds for all algorithms that terminate in a finite number of rounds.
Hence, it is not too surprising that the stronger guarantees provided by the $\ktone$ assumption make it significantly harder to show meaningful lower bounds on the required communication.

While the above-mentioned work assumes the synchronous model, where nodes can send messages of $O(\log n)$ bits over each link in each round, communication complexity has also been studied in the asynchronous $\congest$ $\ktone$ model~\cite{DBLP:conf/wdag/MashreghiK18,DBLP:conf/wdag/MashreghiK19}, which, analogously to its synchronous counterpart, assumes that a node can send a message of size $O(\log n)$ bits whenever it is scheduled to take a step.
Even though time-encoding is not possible in the asynchronous model due to the absence of lock-step synchronicity, the state of the art is that nontrivial lower bounds on the required communication for many graph problems are still lacking even in the asynchronous setting.

In this paper, we introduce a new way of quantifying the communication complexity of distributed graph problems, and our main result is a lower bound on the necessary communication for constructing a spanner with a distributed algorithm.
Given an unweighted graph $G$, a multiplicative \emph{$k$-spanner} \cite{DBLP:journals/jgt/PelegS89,DBLP:journals/jacm/AwerbuchGPV90} is a subgraph $S \subseteq G$, such that the distance in $S$ between any pair of vertices is at most $k$ times their distance in the original graph $G$.
Fast distributed algorithms for constructing a $(2t-1)$-spanner with $O(n^{1+1/t}\log n)$ edges with high probability\footnote{We say that an event happens \emph{with high probability} if it occurs with probability at least $1 - \tfrac{1}{n}$.} are well known, e.g., see \cite{DBLP:journals/talg/ElkinN19}.
Moreover, this size-stretch ratio is believed to be close to optimal due to the girth conjecture of Erd\"os~\cite{erdos}, namely that there exist graphs with $\Omega(n^{1+1/t})$ edges and girth $2t+1$.
We point out that all existing distributed spanner algorithms for the $\congest$ model send at least $\Omega(m)$ bits in the worst case on graphs having $m$ edges.
On the other hand, for the $\local$ model, \cite{DBLP:conf/wdag/BittonEIK19} showed that it is possible to obtain a spanner in constant rounds with $O(n^{1+\epsilon})$ edges and constant stretch by sending only $O(n^{1+\epsilon})$ messages.
Note, however, that there is no bandwidth restriction in the $\local$ model and hence their result does not imply an upper bound on the communication complexity.
Since a spanner can be used as a communication backbone to solve other problems such as single-source broadcast with $o(m)$ bits of communication, we know from the existing lower bound of \cite{jacm15} that, in the clean network model, constructing a spanner is subject to the above-mentioned barrier of $\Omega(m)$ on the message complexity; we note that the $\Omega(m)$ lower bound for spanners was also pointed out in \cite{DBLP:conf/podc/Elkin07}.
In other words, the existing distributed spanner algorithms are near-optimal with respect to time, as well as communication complexity in the clean network model, for small values of $t$.
So far, however, a lower bound on their communication complexity under the $\ktone$ assumption has been elusive. Our work is a first step towards resolving this open question.

\subsection{Our Results} \label{sec:results}

In this work, we initiate a systematic approach to proving lower bounds on the required communication for graph problems under the $\ktone$ assumption.
Our main results are as follows:
\begin{compactitem}

\item
In Section~\ref{sec:lic}, we define the \emph{local information cost} ($\lic$) of solving a problem $P$ with error at most $\gamma$, which we denote by $\lic_\gamma(P)$, and show that it represents a lower bound on the communication complexity in the asynchronous message passing model, as well as a lower bound of $\Omega\lb(\frac{\lic_\gamma(P)}{\log\tau \log n}\rb)$ bits for $\tau$-round algorithms in the synchronous $\congest$ $\ktone$ model.
To the best of our knowledge, this is the first lower bound that provides a general trade-off between communication and time in this setting.
As the local information cost can be characterized for any graph problem $P$, we believe these results to be of independent interest.
Moreover, we show that the local information cost implies a
time complexity lower bound of $\Omega\lb(\frac{\lic_\gamma(P)}{n \log^4 n}\rb)$ rounds in the node-capacitated clique~\cite{DBLP:conf/spaa/AugustineGGHSKL19}, and $\Omega\lb(\frac{\lic_\gamma(P)}{n \log^3 n}\rb)$ rounds in the push-pull gossip model where nodes send messages of size $O(\log n)$.\footnote{Much of the classic literature gossip model does not restrict the message sizes. However, more recent work (e.g. \cite{DBLP:conf/podc/HaeuplerMS18}) considers algorithms in the gossip model where the size of messages is limited to $O(\log n)$ bits.}
\item
We show that constructing a $(2t-1)$-spanner with $O(n^{1+\frac{1}{t} + \epsilon})$ edges, for any $\epsilon \le \frac{1}{64t^2}$, incurs a high local information cost, resulting in a communication complexity of $\Omega\lb(\tfrac{n^{1+1/2t}}{t^2\log n}\rb)$ bits in the $\congest$ $\ktone$ model, for any algorithm that terminates in $O(\poly(n))$ rounds (Theorem~\ref{thm:CONGESTKTONE}), which is the first nontrivial lower bound for a sparse subgraph problem in this setting.
This reveals a sharp contrast to the known fast $(2t-1)$-spanner algorithms such as \cite{baswana}, which takes only $O(t^2)$ rounds and hence does not depend on $n$ at all, but sends $\tilde\Omega(t\cdot m)$ bits.
Interestingly, our lower bound result holds even in the synchronous congested clique, thus showing that the availability of additional communication links does not help in achieving simultaneous time- and message-optimality.
We obtain these results by proving a lower bound on the local information cost for constructing a spanner in the asynchronous $\ktone$ clique. In the proof, we use tools from information theory to quantify the information that many nodes need to learn about their incident edges and then
apply known synchronization techniques.

\item
As a consequence of the above, we obtain Corollary~\ref{cor:simultaneous}, which states that it is impossible to obtain a spanner algorithm that is both time- and communication optimal.

\item
Our technique also implies a \emph{time} complexity lower bound of $\Omega\lb(\tfrac{n^{{1}/{2t}}}{t^2\log^4 n}  \rb)$ rounds for constructing a $(2t-1)$-spanner in the node-capacitated clique of \cite{DBLP:conf/spaa/AugustineGGHSKL19} (see Theorem~\ref{thm:NODECONGESTEDCLIQUE}).
Similarly, we obtain a lower bound of $\Omega\lb(\tfrac{n^{{1}/{2t}}}{t^2\log^3 n}  \rb)$ rounds in the push-pull gossip model with restricted bandwidth (see Theorem~\ref{thm:GOSSIP}).
\end{compactitem}

\subsection{Technical Approach of the Lower Bound for Graph Spanners} \label{sec:technical approach}

To obtain a lower bound on the local information cost of computing a $(2t-1)$-spanner, we consider a graph $G$ on two sets of nodes $U$ and $V$, each of size $n/2$.
The topology of $G$ consists of some static edges and a smaller number of randomly sampled edges.
The static edges are obtained by connecting $U$ and $V$ by many disjoint copies of a complete bipartite graph, which we call the \emph{regions} of $G$.
In contrast, the random edges are independently sampled only on $U$ and loosely interconnect the regions on $U$'s side (see Figure~\ref{fig:lbgraph} on Page~\pageref{fig:lbgraph} where static edges are blue and random edges are red).
The intuition behind this construction is that from the point of view of a node $u \in U$, most of its incident static edges cannot be part of a $(2t-1)$-spanner, whereas most of $u$'s incident random edges {are} \emph{crucial} for maintaining the stretch bound.
That is if a random edge $(u,v)$ is removed, and $u$ and $v$ are not part of the same region of $G$, then it is unlikely that there is a path of length at most $(2t-1)$ from $u$ to $v$, due to the topological structure of $G$.

It is, of course, possible (and even likely when considering existing algorithms such as \cite{baswana}) that some nodes end up including most of their incident edges as part of the spanner.
Nevertheless, most nodes in $U$ need to restrict their output to roughly their (expected) number of incident random edges to guarantee the bound on the size of the spanner. For the rest of this overview, we focus on this set of \emph{sparse nodes}.
While the lower bound graph construction assures us that the crucial random edges must be part of the spanner, we also need to quantify the information that $u$ needs to receive in order to distinguish its static edges from its random ones.
Initially, $u$'s knowledge is limited to its local view of the graph topology, consisting only of its incident edges and the IDs of its neighboring nodes.
Assuming the algorithm terminates correctly, the (initially) high entropy regarding which of $u$'s incident edges are crucial random edges must be reduced significantly over the course of the algorithm.

\subsection{Can We Use 2-Party Communication Complexity?}
It is unclear whether we can obtain a communication complexity bound for spanners under the $\ktone$ assumption via the standard route of two-party communication complexity.
First off, much of the work in this setting (e.g. \cite{DBLP:conf/innovations/FernandezW020}) assumes that the edges are partitioned between Alice and Bob, which makes it challenging to apply these results to the $\ktone$ setting, where edges are always shared between nodes that are neighbors.
On the other hand, if we consider the vertex-partition model, where the vertices and their incident edges are distributed between Alice and Bob (with shared edges being duplicated), it is possible to simulate the $3$-spanner algorithm of \cite{baswana} by using only $O(n\log n)$ bits of communication, no matter how dense the input graph is:
Initially, Alice and Bob locally sample the clusters required by the first phase of the algorithm as follows: They each add all nodes with a degree of at most $\sqrt{n}$ to their respective set of clusters.
Then, they both locally sample each of their remaining vertices independently with probability $\Theta(\log n /\sqrt{n})$.
This ensures that every node that has a degree of at least $\sqrt{n}$ will have a cluster neighbor with high probability.
Alice and Bob then exchange their computed cluster assignments using $O(n\log n)$ bits of communication.
In the second phase of \cite{baswana}, for every vertex, we need to add an edge to each of its neighboring cluster. Since Alice and Bob both know the cluster membership of all nodes, they can simulate this step without further communication by using the simple rule\footnote{A similar rule was used to simulate the algorithm of \cite{baswana} in the LCA model, see \cite{DBLP:conf/innovations/ParterRVY19}.} that, for each vertex $v$ and each of its adjacent cluster $c$, we add a spanner edge to the neighbor of $v$ that has the smallest ID among all neighbors that are members of $c$.

\subsection{Additional Related Work}
Complexity bounds on the message complexity for distributed graph algorithms have been shown for a variety of graph problems under the $\ktzero$ assumption such as computing a maximal independent set~\cite{DBLP:conf/wdag/PaiPPR017} and vertex coloring.
The $\ktone$ assumption was introduced in ~\cite{DBLP:journals/jacm/AwerbuchGPV90}, where the show a message complexity of $\Omega(m)$ for comparison-based broadcast algorithms, where $m$ is the number of edges in the graph.
More recently, \cite{DBLP:conf/podc/PaiPP021}, show several message complexity lower bounds that apply for comparison-based algorithms for symmetry breaking problems, as well as a general lower bound of $\Omega(n)$ message complexity for $O(1)$-ruling sets.  

Numerous hardness results on the time complexity are known in the CONGEST model, the first of which was the work in ~\cite{DBLP:journals/siamcomp/SarmaHKKNPPW12}. 
Here, we focus only on the works that are most closely related to graph spanners.
\cite{censor2018distributed} show a hardness result for the directed k-spanner problem. 
In particular, they show that any $\alpha$-approximation requires at least $\Omega\lt( \frac{\sqrt{n}}{\sqrt{\alpha}\log n} \rt)$ rounds. 
For computing the network diameter, ~\cite{DBLP:conf/soda/FrischknechtHW12} show that a linear number of rounds are necessary. 
While the work of \cite{censor2018distributed,DBLP:conf/soda/FrischknechtHW12} is concerned with \emph{time} complexity, they can be extended to yield message complexity results via the synchronization theorem of  \cite{DBLP:conf/sirocco/PanduranganPS16}.

Several ways of measuring the information learned by the parties when executing an algorithm have been defined in the information and communication complexity literature.
Closely related to our notion of {local information cost} is the \emph{internal information cost} that has been studied extensively in the $2$-player model of communication complexity (see \cite{DBLP:journals/siamcomp/BarakBCR13,chakrabarti2001informational,bar2004information}):
Consider inputs $X$ and $Y$ of Alice and Bob respectively that are sampled according to some joint distribution $\mu$ and let $\Pi$ denote the transcript of a $2$-party protocol.
The \emph{internal information cost with respect to $\mu$} is defined as
$\II[ X : \Pi \mid Y ] + \II[ Y : \Pi \mid X]$.
This corresponds to the expected amount of information that Alice learns about the input of Bob (and vice versa) from the transcript of the protocol.

In contrast, the \emph{external information cost}~\cite{chakrabarti2001informational} measures the amount of information that the transcript reveals to an external observer (without any prior knowledge) about the parties' inputs. In \cite{DBLP:conf/wdag/KolOS17}, it is shown how to generalize the {external information cost} of \cite{DBLP:journals/siamcomp/BarakBCR13} to the multiparty setting in the shared blackboard model, which measures the amount of information that an external observer learns about the players' inputs.
\cite{DBLP:conf/focs/BravermanEOPV13} define the \emph{switched information cost} for the multiparty number-in-hand message passing model, which, in a sense, combines internal and external information cost.
Their model is similar to our setting, with the main difference being that players communicate only through a coordinator. They use the switched information cost to obtain a tight bound of $\Omega(k n)$ on the communication complexity of set disjointness.

\subsection{Roadmap}
In Section~\ref{sec:preliminaries}, we define the distributed computing models and complexity measures. In Section~\ref{sec:lic}, we introduce the local information cost and show how it relates to communication complexity. 
In Section~\ref{sec:lbgraph}, we define the lower bound graph for graph spanners and prove certain reachability properties that are necessary for obtaining a lower bound on the local information cost of constructing spanners in Section~\ref{sec:information}.

\section{Distributed Computing Models}
\label{sec:preliminaries}
For the main technical part of the paper, we consider an {asynchronous network}, where $n$ nodes execute a distributed algorithm and
communicate by sending messages across point-to-point links.
The set of nodes, together with the available communication links, form a clique, and we refer to this model as the \emph{asynchronous clique}.
Since we are interested in studying graph problems in this setting, we consider a graph $G$ with $m$ edges as the input, which is a spanning subgraph of the clique defined above.
Thus, each node of the communication network (i.e.\ clique) is associated with one vertex in $G$ and its incident edges.
We equip the nodes with unique identifiers (IDs) chosen from an integer range of small polynomial size such that an ID can be represented using $\Theta(\log n)$ bits.
Throughout this work, we consider the $\ktone$ assumption introduced by \cite{DBLP:journals/jacm/AwerbuchGPV90}.
In our setting, the $\ktone$ assumption means that each node starts out knowing its own ID in addition to the IDs of all other nodes.
In particular, a node $u$ is aware of the IDs of its neighbors in $G$. We point out that the $\ktone$ assumption has been used by several recent works to obtain sublinear (in the number of edges) bounds on the communication complexity for various graph problems (e.g., \cite{DBLP:journals/jacm/AwerbuchGPV90,DBLP:conf/wdag/GhaffariK18a,DBLP:conf/wdag/GmyrP18,DBLP:conf/podc/KingKT15,DBLP:conf/icdcn/MashreghiK17,DBLP:conf/wdag/MashreghiK18,DBLP:conf/wdag/MashreghiK19}).

Nodes take computing steps at points in time determined by the (adversarial) scheduler, who also controls the speed at which messages travel across links with the restriction that each message sent takes at most one unit of time to be delivered to its destination.
Whenever a node takes a step, it can process all received messages, perform some local computation, including accessing a private source of unbiased random bits, and send (possibly distinct) messages to an arbitrary subset of its peers in the clique.
We assume that each message contains the ID of the sender in addition to the actual payload, and hence the smallest message size is $\Theta(\log n)$ bits.
We say that an algorithm $\cA$ errs with probability at most $\gamma$, if, for any given graph $G$, the execution of $\cA$ has probability at least $1 - \gamma$ to yield a correct output at each node.

In Sections~\ref{sec:conversion} and \ref{sec:applications}, we also consider the \emph{congested clique} \cite{DBLP:journals/dc/LotkerPP06} and the \emph{synchronous $\congest$ $\ktone$ model}~\cite{peleg_book}, where the computation is structured in rounds, and thus each node can send at most $O(\log n)$ bits over each communication link per round.
The former model allows all-to-all communication, analogously to the asynchronous clique described above, whereas the $\congest$ $\ktone$ model restricts the communication to the edges of $G$.

The \emph{message complexity} of a distributed algorithm $\cA$ is defined as the maximum number of messages sent by the nodes when executing $\cA$.
The \emph{communication complexity} of $\cA$, on the other hand, takes into account the maximum number of bits sent in any run of $\cA$.
As these two quantities are within logarithmic factors of each other in the $\congest$ model, we will state our bounds in terms of communication complexity.
When considering the synchronous $\congest$ $\ktone$ model in Section~\ref{sec:applications}, we are interested in algorithms that ensure that all nodes terminate within $\tau$ rounds in the worst case, and we define $\tau$ to be the \emph{time complexity} of such an algorithm.

\section{Local Information Cost}
\label{sec:lic}

When solving a graph optimization problem $P$, each node must produce an output that is a function of its $O(1)$-hop neighborhood. %
In contrast to verification problems such as graph connectivity testing, it is possible that the nodes' output for solving an optimization problem on a given graph is not unique. 
For instance, when constructing a $k$-spanner, the local output of a node is the set of its incident edges that are part of the spanner, with the requirement that the total number of output edges is sufficiently small and that their union guarantees a stretch of at most $k$.

As mentioned in Section~\ref{sec:preliminaries}, we consider an asynchronous clique as the underlying communication network, and we assume that we sample the graph $G$ according to some distribution $\cG$ on a set of $n$ vertices. %
Given $G$, each node $u_i$ observes its initial local state represented by a random variable $X_i$.
At the very least, $X_i$ contains $u_i$'s ID as well as the IDs of its neighbors in $G$, according to the $\ktone$ assumption.
For technical reasons, it sometimes makes sense to reveal additional information about $G$ to $u_i$ (as part of $X_i$); we will do so in Section~\ref{sec:information}.
Clearly, this can only help the algorithm and hence strengthens the obtained lower bound.

In the remainder of the paper, we refer to several standard notions in information theory.
Section~\ref{sec:tools} gives a brief summary of the main definitions and tools.

\begin{definition} \label{def:lic} %
  For an algorithm $\cA$ that errs with probability at most $\gamma$, let random variable $\Pi_i$ be the transcript of the messages received by node $u_i$. %
  We use $\lic_{\cG}(\cA)$ to denote the \emph{local information cost of algorithm $\cA$ under distribution $\cG$}, and define
  \begin{align}
    \lic_{\cG}(\cA)
      = \sum_{i=1}^n
          \II\lb[ \Pi_i : G\ \middle|\ X_i \rb].
    \label{eq:lic_alg}
  \end{align}
  The \emph{local information cost for solving problem $P$ with error at most $\gamma$} is defined as
  \begin{align}
    \lic_\gamma(P) = \inf_{\text{$\cA$: $\gamma$-error}}\
                     \max_\cG\
                     \lic_{\cG}(\cA). \label{eq:lic}
  \end{align}
\end{definition}
That is, we obtain $\lic_{\gamma}(P)$ by considering the best-performing $\gamma$-error algorithm (if one exists\footnote{Note that it may be possible to define an infinite sequence of algorithms with strictly decreasing $\lic$ values.}) and the worst-case input graph distribution $\cG$ with respect to $\cA$.
We point out that similar lower bound arguments based on the concept of mutual information were employed in the context of triangle finding and listing; see \cite{DBLP:conf/spaa/FischerGKO18,DBLP:conf/podc/IzumiG17,DBLP:conf/spaa/Pandurangan0S18}.

Lemma~\ref{lem:compactness} confirms that there always exists a worst-case distribution $\cG$, which justifies using the maximum instead of the supremum in \eqref{eq:lic}.
\newcommand{\lemCompactness}{%
  Consider any $\gamma$-error algorithm $\cA$ as stated in Definition~\ref{def:lic}.
  Then,
  \[
    \max_\cG\lb( \lic_{\cG}(\cA) \rb)
    =
    \sup_\cG\lb( \lic_{\cG}(\cA) \rb).
  \]
}
\begin{lemma} \label{lem:compactness}
  \lemCompactness
\end{lemma}
\begin{proof}
  Let $\cS$ be the set of all graphs of $n$ nodes where each node has a unique ID and let $s = |\cS|$.
  Consider the set $\cD$ of all probability distributions over graphs in $\cS$.
  We will first show that $\cD$ is compact by showing that it is bounded and closed.
  If we represent each element $a \in \cD$ as a vector of length $s$, where entry $a[i]$ is the probability of sampling the $i$-th graph assuming some fixed ordering of $\cS$, it follows that $\cD$ corresponds to a bounded subset of $\mathbb{R}^s$.

  To see that $\cD$ is closed (i.e.\ contains all limit points), consider the map $f : \mathbb{R}^s \ra \mathbb{R}$ where
  \[
    f(a) = \sum_{i=1}^{s} a[i].
  \]
  Since $f$ is continuous and the set $\set{1}$ is closed in $\mathbb{R}$, it follows from Theorem~18.1 in \cite{munkres2013topology} that $f^{-1}(\set{1})$ is closed too.
  Moreover, the hypercube $[0,1]^s$ is closed and hence
  \[
    [0,1]^s \cap f^{-1}(\set{1}) = \cD
  \]
  is closed too.
  Thus we have shown that $\cD$ is compact.
  Now, consider the function $g : \cD \ra \mathbb{R}$ defined as
  $
    g(\cG) = \lic_{\cG}(\cA)
           = \sum_{i=1}^n \II\lb[ \Pi_i : G\ \middle|\ X_i \rb],
  $
  for every $\mathcal{G} \in \mathcal{D}.$

  Since the entropy of discrete random variables is a continuous function, \eqref{eq:mutual2} tells us that the same holds for the term $\II\lb[ \Pi_i : G\ \middle|\ X_i \rb]$, and thus $g$ is continuous too.  
  As we have shown that $\cD$ is compact, Theorem~26.5 in \cite{munkres2013topology} yields that $g(\cD)$ is compact too.
The lemma follows since supremum and maximum coincide in compact spaces.
\end{proof}

We emphasize that, in our definition of $\lic_{\cG}(\cA)$, the node inputs $X_i$ and $X_j$ are not necessarily independent for nodes in $G$;
for instance, if $u_i$ and $u_j$ have a common neighbor, its ID will show up in both $X_i$ and $X_j$.
Consequently, if $u_i$ has neighbors $u_1,\dots,u_j$, then $u_i$'s input $X_i$ is fully determined by its neighbors' inputs $X_1,\dots,X_j$.
This is a crucial feature of the $\ktone$ assumption and a difference to the edge-partition multiparty number-in-hand model of communication complexity (c.f.\ \cite{DBLP:journals/siamcomp/PhillipsVZ16,DBLP:journals/dc/WoodruffZ17}), where the inputs of parties may be chosen to be completely disjoint.

\subsection{Lower Bounds for Distributed Algorithms via Local Information Cost} \label{sec:conversion}

We now prove that the local information cost presents a lower bound on the communication complexity of distributed graph algorithms.
We first show the result for the asynchronous $\ktone$ clique.

\begin{lemma} \label{lem:conversion_async}
  The communication complexity of solving problem $P$ in the asynchronous $\ktone$ model with error at most $\gamma$, is at least $\lic_\gamma(P)$.
\end{lemma}
\begin{proof}
  Define the \emph{communication complexity $\CC(\cA)$ of algorithm $\cA$} to be the maximum number of bits sent in any run of the $\gamma$-error algorithm $\cA$,
  and recall that the \emph{communication complexity $\CC_\gamma(P)$ of solving problem $P$ with error $\gamma$} (e.g., see \cite{ccbook}) is defined  as
  \[
    \CC_\gamma(P) = \min_{\text{$\cA$: $\gamma$-error}} \CC(\cA).
  \]
  Let $\cA$ be a communication-optimal $\gamma$-error algorithm for $P$, i.e., $\CC(\cA) = \CC_\gamma(P)$.
  Moreover, let $G$ be sampled according to $\cG$, which is a distribution that maximizes \eqref{eq:lic} with respect to $\cA$.
  For each node $u_i$, let $L_i$ be the length of the transcript received by $u_i$ during the run of algorithm $\cA$ on graph $G$. Notice that $L_i$ is a random variable and hence different from the \emph{worst case} transcript length, usually denoted by $|\Pi_i|$ in the literature (e.g.\ \cite{DBLP:journals/jcss/Bar-YossefJKS04}).
  We have
  \begin{align}
    \EE[ L_i ]
            \ge  \HH\lb[ \Pi_i  \rb]
            \ge  \HH\lb[ \Pi_i \mid X_i \rb]
            \ge  \II\lb[ \Pi_i : G \mid X_i \rb],
            \label{eq:lic_mutual}
  \end{align}
  where we have used Lemma~\ref{thm:shannon} in the first inequality, \eqref{lem:entropy_conditioning} in the second inequality, and Lemma~\ref{lem:mutual1} in the last step.
  Let random variable $M$ be the number of bits sent by algorithm $\cA$, i.e., $M = \sum_{u_i \in G} L_i$.
  Since $\cA$ is communication-optimal w.r.t.\ $P$, we have %
  \begin{align}
    \CC_\gamma(P)= \CC(\cA) \ge
    \EE[M] &= \sum_{u_i \in G}^{} \EE[L_i] \notag\\
          &\ge \sum_{u_i \in G}^{} \II\lb[ \Pi_i : G \mid X_i \rb]
            \tag{by \eqref{eq:lic_mutual}}\\
          &= \lic_{\cG}(\cA) \tag{by \eqref{eq:lic_alg}}\\
          &= \max_{\cG}\lic_{\cG}(\cA) \tag{by def.\ of $\cG$}\\
          &\ge \inf_{\text{$\cA$: $\gamma$-error}}\
               \max_{\cG} \lic_{\cG}(\cA)  \notag\\
          &= \lic_\gamma(P). \tag{by \eqref{eq:lic}}
  \end{align}
\end{proof}

\begin{lemma} \label{lem:conversion_sync}
  The communication complexity of solving problem $P$ in at most $\tau$ rounds with probability at least $1-\gamma$ is  $\Omega\lb(\frac{\lic_\gamma(P)}{\log \tau \log n}\rb)$ bits.
  This holds in the synchronous $\congest$ $\ktone$ model as well as in the congested clique.
\end{lemma}
\begin{proof}
  Several synchronizers have been proposed in the literature (see \cite{DBLP:journals/jacm/Awerbuch85,peleg_book}).
  As we consider a clique communication topology, we will use the
  $\sigma$-synchronizer of \cite{DBLP:conf/sirocco/PanduranganPS16}.
  Given a synchronous algorithm with communication complexity $C_S$ and round complexity $T_S$ for some problem $P$, the $\sigma$-synchronizer yields an algorithm in the asynchronous clique with a communication complexity of $\CC_\gamma(P) = O(C_S\log(T_S)\log n)$ bits (see Theorem~1 in \cite{DBLP:conf/sirocco/PanduranganPS16}) by exploiting the clique topology and compressing silent rounds.
  Applying Lemma~\ref{lem:conversion_async}, we know that
  \onlyLong{
  \begin{align}
    \lic_{\gamma}(P) &\le \CC_\gamma(P) = O(C_S\log T_S\log n) \notag
  \intertext{and hence}
    C_S &= \Omega\lb(\frac{\lic_\gamma(P)}{\log T_S\log n}\rb). \notag
  \end{align}
  }%
  \onlyShort{
  \[
    \lic_{\gamma}(P) \le \CC_\gamma(P) = O(C_S\log T_S\log n),
  \]
    {and hence}
  $C_S = \Omega\lb(\frac{\lic_\gamma(P)}{\log T_S\log n}\rb).$
  }
  This shows the result for the congested clique.
  Since the $\congest$ $\ktone$ model can be simulated in the congested clique by simply ignoring communication links that are not part of the graph, the same lower bound also holds in the former model.
\end{proof}

We now move on to models where a high local information cost implies a lower bound on the time complexity, due to limitations imposed on the communication capabilities of the nodes.
In more detail, the proofs of the next two lemmas exploit that the total per-round bandwidth is $O(n\polylog n)$ in the node congested clique and the gossip model:

\newcommand{\lemConversionNodeCongested}{
Solving problem $P$ with probability at least $1-\gamma$ in the node-capacitated clique requires $\Omega\lb(\frac{\lic_\gamma(P)}{n \log^4 n}\rb)$ rounds.
}
\begin{lemma} \label{lem:conversion_nodecongested}
  \lemConversionNodeCongested
\end{lemma}
\begin{proof}
  Assume towards a contradiction that there exists an algorithm $\cA$ that solves $P$ (w.h.p.) in $\tau = o\lb(\frac{\lic_\gamma(P)}{n \log^4 n}\rb)$ rounds in the node-capacitated clique.
  During each of the at most $\tau$ rounds of execution, a node can send at most $O(\log n)$ messages of $O(\log n)$ bits according to the specification of the node-capacitated clique~\cite{DBLP:conf/spaa/AugustineGGHSKL19}, i.e., the total number of bits sent per round is $O(n\log^2 n)$.
  Now, suppose we execute the presumed algorithm $\cA$ in the $\ktone$ congested clique.
  Observe that
  $\lic_\gamma(P)
      = \sum_{u_i}^{}\II[ G : \Pi_i \mid X_i ]
      \le n\HH[ G ]
      = \tilde O(n^3)$,
  since each one of the possible $O(n^2)$ edges of $G$ can be described using $O(\log n)$ bits,
  which implies that $\tau = O(\poly(n))$.
  Therefore, the communication complexity $\CC(A)$ of $\cA$ is
  \begin{align}
    \CC(\cA)
    =
    O(n\log^2 n) \cdot \tau
    =
    o \lb( \frac{\lic_\gamma(P)}{\log^2 n} \rb) \notag
    =
    o \lb( \frac{\lic_\gamma(P)}{\log \tau\log n} \rb), %
  \end{align}
  which is a contradiction to Lemma~\ref{lem:conversion_sync}.
\end{proof}

\newcommand{\lemConversionGossip}{
Consider a gossip model as described in \cite{DBLP:conf/stoc/Censor-HillelHKM12,DBLP:conf/podc/HaeuplerMS18} where, in each round, each node can initiate a message exchange with a single neighbor but, in contrast to \cite{DBLP:conf/stoc/Censor-HillelHKM12}, we follow the assumption of \cite{DBLP:conf/podc/HaeuplerMS18} by limiting the bandwidth of each link to $O(\log n)$ bits.
Solving problem $P$ with probability at least $1-\gamma$ requires $\Omega\lb(\frac{\lic_\gamma(P)}{n \log^3 n}\rb)$ rounds.
}
\begin{lemma} \label{lem:conversion_gossip}
  \lemConversionGossip
\end{lemma}
\begin{proof}
  We omit the details as the proof follows along similar lines as the proof of Lemma~\ref{lem:conversion_nodecongested}, with the main difference being that the total available bandwidth that can be used per round is even smaller (by a logarithmic factor) than in the node congested clique.
\end{proof}

\subsection{Limitations of the $\lic$ Approach}

Thus far we have shown that a bound on the $\lic$ yields bounds for several models of interest. 
However, we point out that, in general, a lower bound approach purely based on the $\lic$ cannot yield communication complexity lower bounds that \emph{exceed} the total output size of the problem at hand. 
This follows from the seminal work of \cite{DBLP:conf/soda/ParterY19a,DBLP:conf/podc/ParterY19}, who give a framework for obtaining distributed algorithms that are ``secure'', in the sense that the nodes only learn their respective output while obtaining no information regarding the inputs or outputs of the other nodes.
In the context of computing a graph spanner, this means that each node only learns about its incident spanner edges, whereas for other problems, such as computing a spanning tree, a node may only learn the ID of its parent, which amounts to an output size of $O(n\log n)$ bits (over all nodes). 
An important consequence of the work of \cite{DBLP:conf/soda/ParterY19a,DBLP:conf/podc/ParterY19} is that additional techniques are required for proving lower bounds on the communication complexity of problems with small outputs. 
This includes, in particular, the fundamental problems of single-source broadcast and computing a spanning tree, for which no nontrivial bounds on the communication complexity are known for time-efficient  non-comparison based algorithms under the $\ktone$ assumption, at the time of writing.

\section{A Lower Bound Construction for Spanners} %
\label{sec:lbgraph}

For a given integer $t\ge 2$, we define
\begin{align}
    k = 2t-1 = O\lb(\lb(\frac{\log n}{\log\log n}\rb)^{1/3}\rb). \label{eq:k}
\end{align}
In this section, we describe the lower bound graph distribution $\cG_k$ on which constructing a multiplicative $k$-spanner with a distributed algorithm incurs a high local information cost.

Consider $n$ vertices and divide them into two sets, each of size $n/2$, called $U = \set{ u_1,\dots,u_{n/2}}$ and $V = \set{ v_1, \dots, v_{n/2}}$.
\footnote{To simplify the presentation, we assume that $n/2$, $n^{\frac{2}{k+1}+\frac{1}{4k^2}}$, $\tfrac{1}{2}n^{1-\frac{2}{k+1}-\frac{1}{4k^2}}$, $n^{\frac{1}{k+1}}$ are integers. }
To ensure that each node has a unique ID, we fix some enumeration of the vertices $u_1,\dots,u_{n/2},v_1,\dots,v_{n/2}$ and choose a permutation of $[1,n]$ uniformly at random as the ID assignment.

We classify every edge as either blue or red.
Note that we only introduce this coloring for the purpose of our analysis, i.e., the edge colors are not part of the nodes' input.
We partition the vertices into subgraphs of size $2n^{\frac{2}{k+1}+\frac{1}{4k^2}}$ that we call \emph{regions}.
A region consists of $n^{\frac{2}{k+1}+\frac{1}{4k^2}}$ nodes each from $U$ and $V$ and we form a complete bipartite graph of \emph{blue edges} between the vertices from $U$ and $V$ that lie in the same region.
For instance, the first region consists of the vertices $u_1,\dots,u_{n^{{2}/{(k+1)}+{1}/{4k^2}}}$ and $v_1,\dots,v_{n^{{2}/{(k+1)}+1/4k^2}}$ and the blue edges between them.
Note that each vertex $u_i$ is in exactly one region, denoted by $P(u_i)$.

Next, we add \emph{red edges} by creating a random graph on the subgraph $G[U]$ (induced by the nodes in $U$) according to the Erdős-Rényi model.
That is, we sample each one of the $|U|(|U|-1)$ possible edges independently with probability $\frac{3}{2n^{1-{1}/(k+1)}}$.
Figure~\ref{fig:lbgraph} depicts an example of a lower bound graph instance.

\definecolor{myblue}{RGB}{80,80,160}
\definecolor{blueedges}{RGB}{48,117,246}
\definecolor{mygreen}{RGB}{80,160,80}

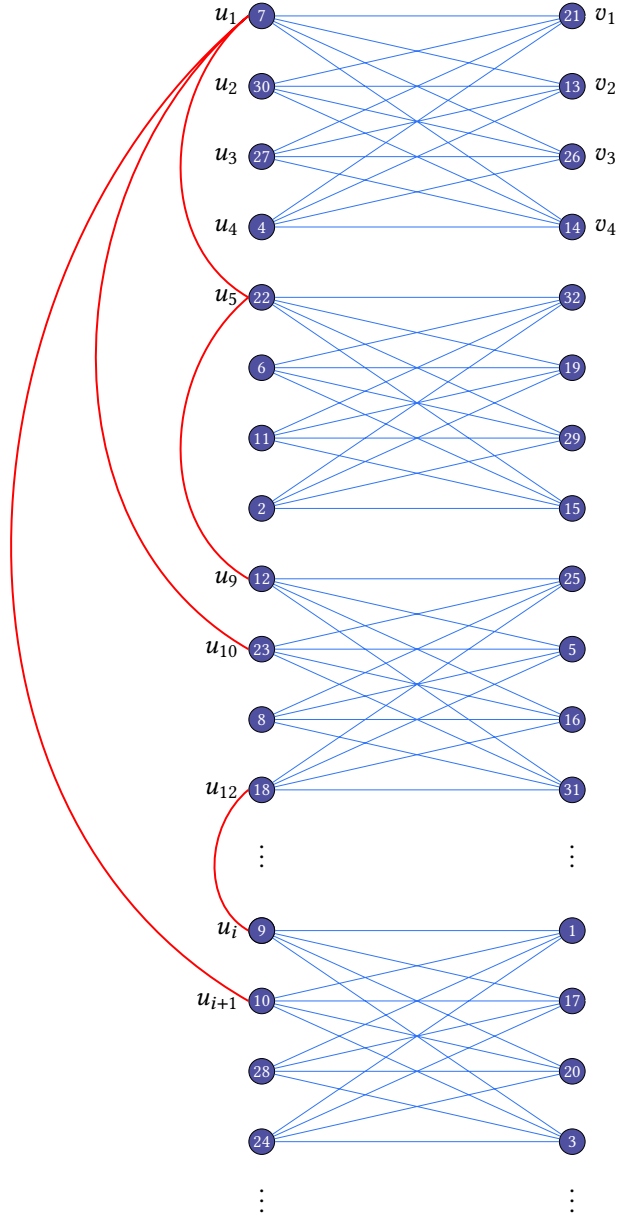
\begin{figure}[h]
\begin{tikzpicture}[scale=0.5]
   \tikzstyle{rededge}=[-,red,thick,auto,out=220,in=150]
   \pgfdeclarelayer{background}
   \pgfdeclarelayer{inbetween}
   \pgfsetlayers{background,inbetween,main}

   \graph[nodes={draw, circle,fill=myblue}, radius=.5cm,
           empty nodes, branch down=1 cm,
           grow right sep=4cm] {
      subgraph I_nm [V={u1, u2, u3, u4}, W={v1,v2,v3,v4}];
      subgraph I_nm [V={u5, u6, u7, u8}, W={v5,v6,v7,v8}];
      subgraph I_nm [V={u9, u10, u11, u12}, W={v9,v10,v11,v12}];
      subgraph[yshift=-1cm] I_nm [V={u13, u14, u15, u16}, W={v13,v14,v15,v16}];

      { u1,u2,u3,u4 } -- [complete bipartite,blueedges] { v1,v2,v3,v4 };
      { u5,u6,u7,u8 } -- [complete bipartite,blueedges] { v5,v6,v7,v8 };
      {u9, u10, u11, u12} -- [complete bipartite,blueedges] {v9,v10,v11,v12};
      {u13, u14, u15, u16} -- [complete bipartite,blueedges] {v13,v14,v15,v16};
    };
      \draw[rededge] (u1.west)  to  (u10.west);
      \draw[rededge] (u1.west)  to  (u14.west);
      \draw[rededge] (u1.west)  to  (u5.west);
      \draw[rededge] (u12.west) to  (u13.west);
      \draw[rededge] (u5.west) to (u9.west);
      \draw[rededge,white,out=320,in=30] (v1.east)  to  (v14.east);

      \node[below,yshift=-0.25cm] at (u12.south) {$\vdots$};
      \node[below,yshift=-0.25cm] at (v12.south) {$\vdots$};
      \node[below,yshift=-0.15cm] at (u16.south) {$\vdots$};
      \node[below,yshift=-0.15cm] at (v16.south) {$\vdots$};
      \node[left] at (u1.west) {$u_1$};
      \node[left] at (u2.west) {$u_2$};
      \node[left] at (u3.west) {$u_3$};
      \node[left] at (u4.west) {$u_4$};
      \node[left] at (u5.west) {$u_5$};
      \node[left] at (u9.west) {$u_9$};
      \node[left] at (u10.west) {$u_{10}$};
      \node[left] at (u12.west) {$u_{12}$};
      \node[left] at (u13.west) {$u_{i}$};
      \node[left] at (u14.west) {$u_{i+1}$};
      \node[right] at (v1.east) {$v_1$};
      \node[right] at (v2.east) {$v_2$};
      \node[right] at (v3.east) {$v_3$};
      \node[right] at (v4.east) {$v_4$};

      \node[white] at (u1) {\tiny $7$};
      \node[white] at (u2) {\tiny $30$};
      \node[white] at (u3) {\tiny $27$};
      \node[white] at (u4) {\tiny $4$};
      \node[white] at (u5) {\tiny $22$};
      \node[white] at (u6) {\tiny $6$};
      \node[white] at (u7) {\tiny $11$};
      \node[white] at (u8) {\tiny $2$};
      \node[white] at (u9) {\tiny $12$};
      \node[white] at (u10) {\tiny $23$};
      \node[white] at (u11) {\tiny $8$};
      \node[white] at (u12) {\tiny $18$};
      \node[white] at (u13) {\tiny $9$};
      \node[white] at (u14) {\tiny $10$};
      \node[white] at (u15) {\tiny $28$};
      \node[white] at (u16) {\tiny $24$};
      \node[white] at (v1) {\tiny $21$};
      \node[white] at (v2) {\tiny $13$};
      \node[white] at (v3) {\tiny $26$};
      \node[white] at (v4) {\tiny $14$};
      \node[white] at (v5) {\tiny $32$};
      \node[white] at (v6) {\tiny $19$};
      \node[white] at (v7) {\tiny $29$};
      \node[white] at (v8) {\tiny $15$};
      \node[white] at (v9) {\tiny $25$};
      \node[white] at (v10) {\tiny $5$};
      \node[white] at (v11) {\tiny $16$};
      \node[white] at (v12) {\tiny $31$};
      \node[white] at (v13) {\tiny $1$};
      \node[white] at (v14) {\tiny $17$};
      \node[white] at (v15) {\tiny $20$};
      \node[white] at (v16) {\tiny $3$};
\end{tikzpicture}
\caption{
  A simplified instance of the lower bound graph sampled according to $\cG_k$ with randomly assigned node IDs.
  The region of $u_1$ consists of nodes $u_1,\dots,u_4,v_1,\dots,v_4$.
  Assuming that $k=5$, the edge $(u_1,u_{i+1})$ is critical as any other path from $u_1$ to $u_{i+1}$ has length greater than $5$.
  On the other hand, the edges $(u_1,u_5)$ and $(u_1,u_{10})$ are not critical as they are both reachable by traversal sequences of length (at most) $k$: after removing $(u_1,u_5$), node $u_5$ is reachable via the traversal sequence $RBBR$.
  Similarly, if we discard $(u_1,u_{10})$, then $u_{10}$ is still in the reachable set $\cR(RRBB)$.
}
\label{fig:lbgraph}
\end{figure}

Let $X$ be the number of red edges incident to $u_i$.
According to the random graph model, we have
\begin{align}
  \EE[X] = (|U|-1)\lb(\frac{3}{2n^{1-{1}/(k+1)}}\rb)
  = \tfrac{3}{4}n^{\frac{1}{k+1}} - o(1). \label{eq:degree_expectation}
\end{align}
The following lemma  shows that all degrees are likely to be concentrated around the mean. We omit the proof, which follows by a straightforward application of a standard Chernoff bound (c.f.\ Theorems~4.4 and 4.5 in \cite{upfalmitzenmacher}).

\newcommand{\lemDense}{
Let $\dense$ be the event that the number of red edges incident to every node in $U$ is
$\frac{3}{4}n^{\frac{1}{k+1}}
  \pm O\lb(n^{\frac{1}{2(k+1)}}\sqrt{\log n}\rb).
$
Then, $\dense$ happens with high probability.
}
\begin{lemma} \label{lem:dense}
  \lemDense
\end{lemma}

\subsection{Reachability and Critical Edges}
\label{sec:reachability}

\paragraph{Critical Edges} Suppose we sample $G$ according to $\cG_k$.
This results in each node $u \in U$ having  $\Theta\lb(n^{\frac{1}{k+1}}\rb)$ incident red edges in expectation.
A crucial property of our lower bound graph is that some of these red edges are likely to be part of any $k$-spanner.
We say that a red edge $(u,v)$ is \emph{critical} if any cycle that contains both $u$ and $v$ has a length of at least $k+2$.
See Figure~\ref{fig:lbgraph} for an example of a critical edge.
From this definition, we immediately have the following property:

\begin{lemma} \label{lem:keepEdges}
  A $k$-spanner of $G$ contains all critical edges.
\end{lemma}

The main result of this section is a lower bound on the expected number of critical edges:

\newcommand{\lemReachability}{%
Conditioned on event $\dense$, every $u \in U$ is incident to at least
$
\tfrac{3}{4}n^{\frac{1}{k+1}}
\lb(1 -
  \frac{2}{n^{{1}/{k^3}}}
\rb)
  - o(1)
$
critical edges in expectation.
}

\begin{lemma} \label{lem:reachability}
  \lemReachability
\end{lemma}

\subsection{Proof of Lemma~\ref{lem:reachability}}

We start our analysis by first defining some technical machinery, after which we give a high-level overview of the proof.

\paragraph{Traversal Sequences and Reachable Sets} \label{sec:traversal}
Consider some node $u$ and an incident red edge $(u,v)$ in $G$, i.e., $v \in U$, and let $H$ be the graph obtained by removing $(u,v)$.
There is a nonzero probability that $v$ is reachable from $u$ in $H$ by traversing some sequence of red and blue edges.
This motivates us to consider a {traversal sequence} (starting from $u$), which is a $k$-length sequence of edge colors that specify, for each step, whether we follow the red or blue edges.
Formally, a \emph{traversal sequence $T$} is a $k$-length string where each character is chosen from the alphabet $\set{R,B}$; $R$ and $B$ stand for colors red and blue, respectively. 
We use the notation $T[i]$ to refer to the $i$-th step of $T$, and define $T[0]=\perp$ for convenience.
Any sequence of $R$ and $B$ can form a traversal sequence, under the restriction that there exists a path in $H$ starting from $u$ that traverses an edge of color $T[i]$ in its $i$-th step, for all $1 \le i \le k$.

We use the notation $T[i,j]$ to denote the \emph{traversal subsequence} $T[i]\ T[i+1]\ \dots T[j]$.
We say that the $i$-th step is a \emph{$B$-step} if $T[i]=B$, and define \emph{$R$-step} similarly. %
The following is immediate from the definition of the lower bound graph:

\begin{lemma} \label{obs:traversal}
Consider any traversal sequence $T$ of length $k \ge 3$ and suppose that $i<k$. 
If $T[i] = B$, then either $T[i-1]=B$ or $T[i+1]=B$ (or both).
\end{lemma} 

A traversal sequence $T$ induces a \emph{reachable set} of nodes, denoted by $\mathcal{R}(T)$, which is determined by the subset of all nodes in $U$ that lie in the $k$-hop neighborhood of $u$ and are reachable from $u$ by following all possible paths in graph $H$ along edges with the colors in the order specified by $T$.
We use $\mathcal{R}_i(T)$ to refer to the reachable set of the traversal subsequence $T[1,i]$.
The following lemma is immediate from the definition of traversal sequences:

\begin{lemma} \label{lem:reachability_tree}
  Node $v\in U$ is reachable from $u\in U$ in at most $k$ hops in $H$ if there exists a traversal sequence $T$ of length $k$ starting at $u$ such that $v \in \mathcal{R}(T)$.
\end{lemma}

\paragraph{High-level Overview of the Proof}
Lemma~\ref{lem:reachability_tree} provides us with the following strategy for bounding the probability that $v$ is reachable from $u$ in $k$ steps:
Consider some $u \in U$ and a red edge $(u,v)$ to some other $v \in U$.
Recall that, if $(u,v)$ is critical then, upon removing $(u,v)$, reaching $v$ from $u$ requires more than $k$ hops.
Thus, we analyze the reachability of $v$ from $u$ on the graph $H = G \setminus \{(u,v)\}$, and assume that event $\dense$ occurs, i.e., the number of incident red edges is close to the expected value for all nodes in $U$.
We first identify the type of traversal sequence that exhibits the largest growth with respect to the reachable set of nodes.
Below we show that sequences that follow a certain structure, i.e., are ``$BB$-maximal'', dominate all other sequences regarding the probability of leading to a reachable set that contains $v$.
Intuitively speaking, a sequence is $BB$-maximal if it alternates crossing the blue edges twice (this will extend the reachable set to all nodes in $U$ that are inside the regions of the currently reachable nodes) with a single hop over red edges (which may lead to nodes outside of theses already-reached regions).
We show that the number of nodes in $U$ reached by a $BB$-maximal traversal sequence is still sufficiently small, which yields an upper bound of roughly $\frac{1}{2^k n^{{1}/{k^3}}}$ on the probability of reaching $v$ with a $BB$-maximal traversal sequence.
This shows that, at least a $\lb(1 - \frac{2}{n^{{1}/{k^3}}} \rb)$-fraction of the incident red edges of $u$ are critical in expectation.
In the remainder of this section, we give the detailed proof.

As it is difficult to directly analyze the size of the reachable set for a given traversal sequence, we instead study the growth of a function $\rho_i(T)$ that, as shown below, upper-bounds the size of this set after the first $i$ steps of $T$.

\begin{definition} \label{def:growth}
Let $T$ be a $k$-length traversal sequence, for some integer $k \ge 3$, and define $T[0]=\perp$.
For each $i \in [0,k]$, we define a function $\rho_i$ where
\begin{subnumcases}
{\rho_{i}(T) :=}
  1 & \text{if $i=0$;} \label{eq:rho1}   \\
  \rho_{i-1}(T)\cdot n^{\frac{1}{k+1}}  & \text{if $i\ge 1$, $T[i]=R$;} \label{eq:rho2}\\
  \rho_{i-1}(T)\cdot n^{\frac{2}{k+1}+\frac{1}{4k^2}} & \text{if $i\ge2$, $T[i-2,i]=XBB$, where $X \ne B$;}  \label{eq:rho4}     \\
  \rho_{i-1}(T) & 
  \text{otherwise.}
  \label{eq:rho3}
\end{subnumcases} 
Moreover, we define $\rho(T) := \rho_k(T)$.
\end{definition}

\begin{lemma} \label{lem:growth}
  Consider any $k$-length traversal sequence $T$ and suppose that event $\dense$ happens (see Lemma~\ref{lem:dense}). 
  Then, it holds that
  $|\mathcal{R}_i(T)| \le \rho_i(T)$, for all $i \in [0,k]$.
\end{lemma}
\begin{proof}
We proceed by induction over $i$.
The base step $i=0$ is trivial since $\mathcal{R}_0(T) = \set{ u }$.
 
For the inductive step, first consider the case $i\ge 1$ and $T[i]=R$, i.e., we traverse the red edges incident to nodes in $\mathcal{R}_{i-1}(T)$.
Since we assume that event $\dense$ occurs, the number of red edges incident to each node in $|\mathcal{R}_{i-1}(T)|$ is bounded from above by $n^{\frac{1}{k+1}}$, and the claim follows.
Next, consider the case where $T[i]=B$ and $T[i-1]\ne B$. 
If $T[i-1]=\perp$ (i.e., $i=1$), then it is clear that we only reach nodes in $V$ by following the blue edges incident to $u$, which implies $\mathcal{R}_1(T) = \mathcal{R}_{i-1}(T)$, and the claim follows by the inductive hypothesis.
Similarly, if $T[i-1]=R$, then we know that the last step used edges connecting nodes in $U$ and hence the $i$-th step of $T$ reaches only nodes in $V$ and thus $\mathcal{R}_1(T) = \mathcal{R}_{i-1}(T)$. 
Again, the statement follows from the inductive hypothesis.

Finally, consider the case $i \ge 2$, $T[i]=B$, $T[i-1]=B$, and $T[i-2] \ne B$.
We pessimistically assume that all nodes in $\mathcal{R}_{i-2}(T)$ lie in $|\mathcal{R}_{i-2}(T)|$ distinct regions of the graph, and let $S$ denote the set of nodes in these regions. 
Recalling that each region forms a complete bipartite graph on $2 n^{\frac{2}{k+1}+\frac{1}{4k^2}}$, it follows that the traversal sequence reaches all nodes in $S \cap V$ in step $i-1$.
Thus, in step $i$, it reaches all nodes in $S \cap U$. 
Since $|S \cap U| = n^{\frac{2}{k+1}+\frac{1}{4k^2}}$, we have
\begin{align}
  |\mathcal{R}_{i}(T)| \le 
  |\mathcal{R}_{i-2}(T)|n^{\frac{2}{k+1}+\frac{1}{4k^2}} 
  &\le \rho_{i-2}(T)\cdot n^{\frac{2}{k+1}+\frac{1}{4k^2}} \notag\\ 
  &= \rho_{i-1}(T) n^{\frac{2}{k+1}+\frac{1}{4k^2}} \tag{by \eqref{eq:rho3}}\\ 
  &= \rho_i(T) \tag{by \eqref{eq:rho4}}
\end{align}
\end{proof}

Throughout this section, whenever we say ``traversal sequence'', we mean a traversal sequence of length $k\ge 3$, for some integer $k$.
Moreover, we refer to two adjacent $B$-steps as a \emph{$BB$-pair}.

\begin{definition} \label{def:bmaximal} 
  A traversal sequence $T$ is \emph{$BB$-maximal}, if all of the following hold:
  \begin{compactenum}
    \item $T$ does not contain three consecutive $B$-steps.
    \item $T$ contains at most one pair of consecutive $R$-steps and, if it does, then this pair occurs at the very end, i.e., $T[k-1]=T[k]=R$.
  \end{compactenum}
\end{definition}
Intuitively speaking, a $BB$-maximal traversal sequence contains the maximal number of $BB$-pairs.
Equations~\eqref{eq:t1} and \eqref{eq:t2} show examples of such a sequence.

Next, we will prove that, even if we expand the reachability set by following a $BB$-maximal traversal sequence, the resulting size still falls short of containing a constant fraction of the nodes in $U$.

\begin{lemma} \label{lem:kmod}
  Condition on event $\dense$.
  If $T$ is $BB$-maximal, then it holds that
  \[
    \rho(T) \le \frac{n^{1-\frac{1}{k^3}}}{2^{k}}.
  \]
\end{lemma}
\begin{proof}
  We distinguish three cases depending on the value of $k\bmod 3$.

  First, consider the case where $k\bmod 3 =0$.
  Let $T$ be a $B$-maximal traversal sequence.
  Definition~\ref{def:bmaximal} tells us that the remaining sequence is fully determined by whether the first character is $B$ or $R$.
  More specifically, the only two possibilities for $T$ are either
  \begin{align}
    T &= \underbrace{BBR\ BBR\ \dots BBR}_{\text{$k/3$-times}}\label{eq:t1}\\ 
    \intertext{or}
    T &= \underbrace{RBB\ RBB\ \dots RBB}_{\text{$k/3$-times}}.\label{eq:t2}
  \end{align}
  Thus, we have exactly $k/3$ triples of the form $BBR$ or $RBB$ in $T$.
  Applying Definition~\ref{def:growth} for each step, we obtain %
  \begin{align}
    \rho(T) = n^{\frac{k}{3} \lb( \frac{2}{k+1} + \frac{1}{2k^2} + \frac{1}{k+1} \rb)}
    =
    n^{\frac{k}{k+1}+\frac{1}{6k}}. \label{eq:RTL}
  \end{align}
  Observe that
  \begin{align}
 n^{1 - \frac{1}{k^3} - \frac{1}{k^2}}
    \le
    n^{1 - \frac{1}{k^3} - \frac{k}{\log_2 n}} 
    = 
    \frac{n^{1-\frac{1}{k^3}}}{2^{k}} 
    \label{eq:RTR}
  \end{align}
  where the inequality follows from \eqref{eq:k}.
  Thus, our goal is to show that the right-hand side of \eqref{eq:RTL} is bounded by the left-hand side of \eqref{eq:RTR}.
  Comparing the exponents in \eqref{eq:RTL} and \eqref{eq:RTR}, we observe that $\rho(T) \le n^{1- \frac{1}{k^3}-\frac{1}{k^2}}$ if
  \begin{align}
    \frac{k}{k+1}+\frac{1}{6k}
    &\le 1 - \frac{1}{k^3} - \frac{1}{k^2},
    \notag
  \end{align}
  Multiplying by the common denominator, this is true if
  \begin{align}
  \frac{k^3 - \frac{7k^2}{5} - \frac{12k}{5} - \frac{6}{5}}{k^3(k+1)} \ge 0.
    \label{eq:kmod0}
  \end{align}
  Consider the function
  $f(k) = k^3 - \frac{7k^2 + 12k + 6}{5}.$
    Factorizing and simplifying shows that $f(k) \ge 0$ for all $k \ge 3$, which implies \eqref{eq:kmod0}.

Next, we consider the case $k\bmod 3 = 1$. 
Since $T$ is $BB$-maximal, there must be exactly $\ell = \frac{k-1}{3}$ $BB$-pairs and $R$-steps in the $(k-1)$-length subsequence $T' = T[1,k-1]$.
First suppose that $T'$ satisfies \eqref{eq:t1}, which means that $T[k] = B$.
In this case, Definition~\ref{def:growth} tells us that $\rho_{k-1}(T)=\rho_{k}(T)$ and hence the result follows from the case $k\bmod 3 = 0$.
Now, suppose that  $T'$ satisfies \eqref{eq:t2}, which means that $T[k] = R$.
Thus, according to Definition~\ref{def:growth}, we get
\begin{align}
  \rho(T') = n^{\frac{k-1}{3} \lt( \frac{2}{k+1} + \frac{1}{4k^2}  + \frac{1}{k+1} \rt) + \frac{1}{k+1}}
  = n^{\frac{k}{k+1} + \frac{1}{12k} - \frac{1}{12k^2}}.
\end{align}
By a calculation similar to the one for the case $k\bmod 3 = 0$, it follows that, for all $k \ge 3$,  
\begin{align}
n^{\frac{k}{k+1} + \frac{1}{12k}  - \frac{1}{12k^2}}
  \le
  n^{1 - \frac{1}{k^3} - \frac{1}{k^2}}
  \le 
  \frac{n^{1 - \frac{1}{k^3}}}{2^k},
\end{align}
where the second inequality follows from \eqref{eq:RTR}.

  The final case that we need to analyze is $k\bmod 3 = 2$.
  Consider the length $(k-2)$ subsequence $T'$ of $T$.
  Since $(k-2)\bmod 3 = 0$, we first consider the case where $T'$ satisfies \eqref{eq:t1}. 
  Since $T$ is $BB$-maximal, it follows that $T[k-1]=T[k]=B$, and therefore $T$ consists of $\frac{k+1}{3}$ pairs of $BB$-steps, interleaved with $\frac{k-2}{3}$ $R$-steps.
  Analogously to the first case, Definition~\ref{def:growth} yields
  \begin{align}
    \rho(T) 
    &=
    n^{\frac{k+1}{3} \lt( \frac{2}{k+1} + \frac{1}{4k^2} \rt) + \frac{k-2}{3} \frac{1}{k+1}}  \\
    &=
    n^{\frac{2}{3} + \frac{1}{6k} + \frac{1}{6k^2} 
       + \lt(\frac{1}{3} - \frac{1}{k+1}\rt)}  \\
    &=
    n^{1 + \frac{1}{6k} + \frac{1}{6k^2} 
       - \frac{1}{k+1}}. \label{eq:third}
  \end{align}
  For any $k\ge 3$, it holds that
  \begin{align}
      1 + \frac{1}{6k} + \frac{1}{6k^2} 
       - \frac{1}{k+1}
      \le
      1 - \frac{1}{k^3} - \frac{1}{k^2},
  \end{align}
  and thus the right-hand side of \eqref{eq:third} is at most $n^{1 -  \frac{1}{k^3} - \frac{1}{k^2}}$ as required.
  
  Now consider the case where the subsequence $T'$ satisfies \eqref{eq:t2}.
  Since $T$ is $BB$-maximal, it holds that $T[k-1]=R$.
  We distinguish two cases depending on the value of $T[k]$:

  \begin{enumerate}
  \item  First, suppose that $T[k]=B$.
  As $T[k-1]=R$ and all other $B$-steps happening before step $k$ must have occurred in pairs, it follows that no new nodes in $U$ are reached in the $k$-th step of $T$.
  Consequently, $\rho(T) = \rho_{k-1}(T)$ and since $T[1,k-1]$ satisfies $k\bmod 3 = 1$, the claim follows from the previously analyzed case.
  
  \item Finally, suppose that $T[k] = T[k-1] = R$.
  Since we have $\frac{k-2}{3}$ $BB$-pairs and $\frac{k-2}{3}+2 = \frac{k+4}{3}$ $R$-steps in $T$, Definition~\ref{def:growth} yields
  \begin{align} 
    \rho(T)
    &=
    n^{\frac{k-2}{3} \lt( \frac{2}{k+1} + \frac{1}{4k^{2}} \rt) 
    +
    \frac{k+4}{3}\frac{1}{k+1}}  \notag\\
    &=
    n^{\frac{2}{3}\lt( 1 - \frac{3}{k+1} \rt) + \frac{1}{12k} - \frac{1}{6k^2} + \lt( \frac{1}{3} + \frac{1}{k+1} \rt)} \notag\\
    &=
    n^{1 - \frac{2}{k+1} + \frac{1}{12k} - \frac{1}{6k^2}}. \notag
  \end{align}
  Recalling \eqref{eq:RTR}, it needs to hold for the exponent on the right-hand side that
  \begin{align} 
    1 - \frac{2}{k+1} + \frac{1}{12k} - \frac{1}{6k^2} 
    \le 
    1 - \frac{1}{k^3} - \frac{1}{k^2}. \notag
  \end{align}
  This is satisfied if
  \begin{align}
  \frac{k^{3} - \frac{k^2}{23} - \frac{12k}{23} - \frac{12}{23}}{k^3(k+1)} \ge 0, \notag
  \end{align}
  and factorizing shows that this holds for any $k \ge 3$, as required.
  \end{enumerate}
 \end{proof}
 
Consider an arbitrary traversal sequence $T$. 
We say that an $R$-step at index $i$ is \emph{free in $T$} if $T[i-1]=R$ or $T[i+1]=R$, or both.
By left-shifting $T[i+1],\dots,T[k]$ by one position and setting $T[k]=R$, we obtain a new traversal sequence $T'$.
We call this operation an \emph{$R$-shift of $T$}.
The following lemma is immediate from Definition~\ref{def:growth}:

\begin{lemma} \label{lem:free}
Suppose we obtain $T'$ by performing an $R$-shift on some traversal sequence $T$. Then, $\rho(T') = \rho(T)$.
Moreover, if $T$ does not contain three consecutive $B$-steps (Property~(1) of Definition~\ref{def:bmaximal} holds), then the same is true for $T'$. 
\end{lemma}

The next Lemma confirms the intuition that a traversal sequence does not reach as many nodes as possible, if it contains a subsequence of $3$ or more consecutive $B$-steps.

\begin{lemma} \label{lem:two_Bs}
  Consider any traversal sequence $T'$ that contains at least three consecutive $B$-steps, i.e., there exists an index $i$ and an integer $\ell \ge 2$, such that $T'[i]=\cdots=T[i+\ell]=B$, $T'[i+\ell+1]=R$; assume that $i$ is the smallest index for which this holds.
  Define traversal sequence $T$, where
  \[
  T[j] =
    \begin{cases}
      R      & \text{if $j \in [i+2,i+\ell]$;} \\
      T'[j]   & \text{otherwise.} \\
    \end{cases}
  \]
  Then, it holds that $\rho(T') < \rho(T)$.
\end{lemma}
\begin{proof}
Recalling that $i$ was chosen to be the smallest index for which the premise of the lemma holds, it follows that either $i=1$ or $T'[i-1]=R$.
By assumption $\rho_{i+1}(T') = \rho_{i+1}(T)$. 
Equation~\eqref{eq:rho2} implies that 
\[
\rho_{i+2}(T) = \rho_{i+1}(T) \cdot n^{1/(k+1)} = \rho_{i+1}(T') \cdot n^{1/(k+1)},
\]
whereas, for $T$, we apply \eqref{eq:rho3}, which ensures that $\rho_{i+2}(T') = \rho_{i+1}(T')$, and thus $\rho_{i+2}(T) > \rho_{i+2}(T')$.
This captures the intuition that, in step $i+2$ in $T$, we traverse red edges, whereas in $T'$, we again take a step along the blue edges, which must lead to nodes that are in regions already contained in the reachable set and hence cannot increase the reachable set of nodes. 
We continue to apply this argument to all indices $i+2,\dots,i+\ell$ and conclude that $\rho_{i+\ell}(T) > \rho_{i+\ell}(T')$.
Since the color sequences of $T$ and $T'$ are the same in the remaining indices $i+\ell+1,\dots,k$, Definition~\ref{def:growth}  tells us that $\rho_{i+\ell+j}(T) > \rho_{i+\ell+j}(T')$, for all $j \in [1,k-\ell -i ]$, and the lemma follows.
\end{proof}

\begin{lemma} \label{lem:two_Rs}
Let $T$ be any traversal sequence that contains at least one $BB$-pair but no three consecutive $B$-steps. 
Let $i$ be the smallest index of $T$ such that $T[i]=T[i+1]=B$.
Consider the traversal sequence $T'$ defined as $T'[i] = T'[i+1] = R$ whereas, for all $j \ne \set{i,i+1}$ we have $T'[j] = T[j]$. 
Then, $\rho(T') < \rho(T)$.
\end{lemma}
\begin{proof}
Since $T$ and $T'$ are identical up to including index $i-1$, we have
$\rho_{i-1}(T) = \rho_{i-1}(T')$.
Then, after following only red edges for two consecutive steps, as required by $T'$, Definition~\ref{def:growth} implies that
\[
\rho_{i+1}(T') = \rho_{i-1}(T)\cdot \lb(n^{\frac{1}{k+1}}\rb)^2 = \rho_{i-1}(T)\cdot n^{\frac{2}{k+1}}.
\]
On the other hand, if we consider $T$ and take two hops following only blue edges, it again follows from Definition~\ref{def:growth} that  
\[
\rho_{i+1}(T) = \rho_{i-1}(T)\cdot n^{\frac{2}{k+1}+\frac{1}{4k^2}},
\]
and hence $\rho_{i+1}(T) > \rho_{i+1}(T')$.
 
To see that this inequality continues to hold for the remaining indices, we use the fact that $T[i+2]=T'[i+2]=R$ and $T$ and $T'$ perform the same sequence of colors for all indices $j \ge i+3$ onward.
\end{proof}

\begin{lemma} \label{lem:bbmax_is_max}
If $T$ is not $BB$-maximal, then there exists a $BB$-maximal traversal sequence $\bar{T}$ such that $\rho(\bar{T}) \ge \rho(T)$.
\end{lemma}
\begin{proof}
In the proof, we combine Lemmas~\ref{lem:free}, \ref{lem:two_Bs}, and \ref{lem:two_Rs}:
Assume towards a contradiction that $T$ is not $BB$-maximal, and $\rho(T) > \rho(\bar{T})$, for any $BB$-maximal traversal sequence $\bar{T}$.
We choose $T$ such that it has the maximum $\rho(T)$ of all non-$BB$-maximal traversal sequences with this property, which implies that $\rho(T) \ge \rho(T')$, for any (not necessarily $BB$-maximal) traversal sequence $T'$.
Since $T$ is not $BB$-maximal, it does not satisfy at least one of the two properties of Definition~\ref{def:bmaximal}.

Suppose that $T$ does not satisfy (1), i.e., it contains a segment of three or more consecutive $B$-steps. 
By applying Lemma~\ref{lem:two_Bs} to the first of these segments, we obtain a distinct traversal sequence $T'$ that satisfies (1) and $\rho(T') > \rho(T)$, which is a contradiction. 

Now, consider the case that $T$ does not satisfy (2), i.e., $T[i]=T[i+1]=R$, for some $i \le k-2$, and suppose that $i$ is the smallest index for which this is true.
First suppose that $T$ does not contain any $BB$-pair after step $i$.
Since $i \le k-2$, it follows that $T$ ends with at least three $R$-steps.
Applying Lemma~\ref{lem:two_Rs} tells us that replacing the last two steps of $T$ with a $BB$-pair yields a traversal sequence $T'$ such that $\rho(T') > \rho(T)$, which is a contradiction to the assumed maximality of $\rho(T)$.

Thus, we can assume that $T$ contains at least one $BB$-pair after step $i$. 
By construction, $T$ has a free $R$-step at index $i$, and thus Lemma~\ref{lem:free} yields a sequence $T_1$ such that $\rho(T) = \rho(T_1)$ by moving this $R$-step to the end and shifting all subsequent steps to the left.
Then, we check if $T_1$ is $BB$-maximal, i.e., satisfies Property~(2) of Definition~\ref{def:bmaximal}, or ends with at least three $R$-steps.
In the former case, we have a contradiction to the assumption that $\rho(T) > \rho(\hat{T})$, for any $BB$-maximal traversal sequence $\hat{T}$, whereas in the latter case we obtain a contradiction by applying Lemma~\ref{lem:two_Rs}.

Otherwise, we continue this process by considering the next smallest index $i'$ for which we can find a free $R$-step in $T_1$, which must exist since $T_1$ is not $BB$-maximal, and obtain sequence $T_2$ with $\rho(T_2) = \rho(T)$, by applying the left-shifting operation of Lemma~\ref{lem:free} to index $i'$.
After at most three iterations of this process, we must arrive at some sequence $T_3$ that is either $BB$-maximal or ends with three $R$-steps.
Either case provides a contradiction to the assumed maximality of $\rho(T)$.
\end{proof}

Combining Lemmas~\ref{lem:kmod} and \ref{lem:bbmax_is_max} yields the following:

\begin{corollary} \label{cor:max} 
For any traversal sequence $T$, it holds that 
\begin{align}
  |\mathcal{R}(T)| \le \frac{n^{1 - \frac{1}{k^3}}}{2^{k}}. \label{eq:RT}
\end{align}
\end{corollary}

We can now complete the proof of Lemma~\ref{lem:reachability}.
Recall that we have analyzed the reachability sets in the graph $H$, which does not contain the edge $(u,v)$.
Moreover, $(u,v)$ is sampled independently of the other red edges according to our lower bound graph distribution $\cG_k$ (see Sec.~\ref{sec:lbgraph}).
It follows that node $v$ is independent of $\mathcal{R}(T)$ and hence the probability that $v \in \mathcal{R}(T)$ is at most $\frac{|\cR(T)|}{|U|} = \frac{|\cR(T)|}{n/2}$.
By taking a union bound over the (at most) $2^k$ possible traversal sequences and using \eqref{eq:RT}, we obtain that
\begin{align}
  \Pr\lb[ \text{$(u,v)$ not critical} \rb]
    = \Pr\lb[ \exists T\colon v \in \mathcal{R}(T) \rb]
    \le \frac{2^{k+1} |\mathcal{R}(T)|}{n}
    \le 2n^{-\frac{1}{k^3}}.\label{eq:critical_bnd}
\end{align}
Let $X$ be the number of red edges incident to $u$, and let $Y \subseteq X$ be the random variable denoting the number of critical edges incident to $u$.
From \eqref{eq:critical_bnd}, we get
\[
  \EE\lb[ Y \ \middle|\ \dense, X \rb]
  \ge X\lb(1 - 2n^{-\frac{1}{k^3}}\rb).
\]

Therefore,
\begin{align}
  \EE\lb[ Y \ \middle|\ \dense \rb]
  &=
  \EE_X\lb[ \EE_Y\lb[ Y \ \middle|\ \dense, X \rb]
        \ \middle|\ \dense
     \rb] \notag\\
  &\ge
  \lb(1 - 2n^{-\frac{1}{k^3}}\rb) \EE \lb[ X
        \ \middle|\ \dense
    \rb] \notag\\
  &\ge
  \lb(1 - 2n^{-\frac{1}{k^3}}\rb)
  \lb(\tfrac{3}{4}n^{\frac{1}{k+1}} - o(1)\rb)
  \tag{by \eqref{eq:degree_expectation}} \\
  &\ge
  \tfrac{3}{4}
  \lb(n^{\frac{1}{k+1}}
    - 2n^{\frac{1}{k+1} - \frac{1}{k^3}}
  \rb)
    - o(1). \notag 
\end{align}
This completes the proof of Lemma~\ref{lem:reachability}.

\section{The Local Information Cost of Spanners}
\label{sec:information}

In this section, we use an information-theoretic approach to bound the local information cost of computing a spanner.
In more detail, we will prove the following result:

\begin{theorem} \label{thm:lic_spanner}
    Consider the problem of constructing a $(2t-1)$-spanner that, with high probability, has at most $O\lb(n^{1+\frac{1}{t}+\frac{1}{16(2t-1)^2}}\rb)$ edges, for any integer $t$ such that $2 \le t = O\lb(\lb(\log n /\log\log n\rb)^{1/3}\rb)$.
    Then,
    \[
      \lic_{\text{\tiny${1}/{n}$}}(\text{$(2t-1)$-spanner}) =
      \Omega\lb(\tfrac{1}{t^2} n^{1+\frac{1}{2t}} \log n\rb).
    \]
\end{theorem}
To put Theorem~\ref{thm:lic_spanner} into perspective, note that the distributed $(2t-1)$-spanner algorithm of Baswana \& Sen~\cite{baswana} outputs at most
$O(t n^{1+\frac{1}{t}}\log n) = O(n^{1+\frac{1}{t}+\frac{t}{\log n}+\frac{\log\log n}{\log n}})$ edges with high probability.
For $t = o\lb( \sqrt{\log n /\log\log n}\rb)$,
this amounts to strictly less than $n^{1+\frac{1}{t}+\frac{1}{16(2t-1)^2} }$ edges.

Consider any $(2t-1)$-spanner algorithm $\cA$ that achieves the properties stated in the theorem's premise.
Using \eqref{eq:k}, we can rewrite the bound on the number of edges in Theorem~\ref{thm:lic_spanner} as $O\lb(n^{1 + \frac{2}{k+1} +\frac{1}{16k^2}}\rb)$, which we will use throughout this section.

We make use of the following indicator random variables:
\begin{compactitem}
  \item We define $I_D \!=\! 1$  if and only if every node $u \in U$ has roughly $\frac{3}{4}n^{\frac{1}{k+1}}$ incident red edges, i.e., event $\dense$ occurs, which happens with high probability according to Lemma~\ref{lem:dense}.

  \item Let $Y_i$ be the number of critical edges incident to $u_i$.  We define $I_{Y_i} \!=\! 1$ if and only if ${Y_i} \ge \tfrac{1}{16}n^{\frac{1}{k+1}}$. %

  \item In parts of our analysis, in particular Lemma~\ref{lem:final}, we will focus on the nodes where the output is sparse, in the sense that each of them adds only few incident edges to the spanner.
  In particular, we will prove that these sparse nodes learn a significant amount of information about which of their incident edges are critical.
  This motivates us to define $\sparse \subseteq U$ to be the set of \emph{sparse nodes}, where every node $u \in \sparse$ outputs at most $n^{\frac{2}{k+1}+ \frac{1}{8k^2} }$ edges.
  We use the indicator random variable $I_{S_i}$ and define $I_{S_i}=1$ if and only if $u_i \in \sparse$ and $|\sparse| \ge \tfrac{n}{2} - \tfrac{n}{\log n}$.
  The reason for introducing $I_{S_i}$ will become clearer in the proof of Lemma~\ref{lem:initial}.
\end{compactitem}
For the indicator random variables defined above, we use shorthands such as $I_{{D,{Y_i}}} \!=\! 1$ to refer to the event $I_D \!=\! I_{Y_i} \!=\! 1$.
\medskip

\paragraph{The Initial Knowledge of Nodes}
We consider the following information to be part of $u_i$'s input $X_i$:
Due to the $\ktone$ assumption, $u_i$ knows the random variable $Z_i$, which contains the list of IDs of its neighbors in $G$ including its own ID. Notice that $Z_i$ does not contain any other information about the network at large.
In addition, node $u_i$ also knows the number of incident critical edges to $u_i$, which we denote by the random variable ${Y_i}$.
This extra knowledge is not part of the $\ktone$ assumption but given to $u_i$ for free.
Since ${Y_i}$ fully determines the indicator random variable $I_{Y_i}$ defined above, it follows that $u_i$ also knows $I_{Y_i}$.
Finally, we assume that $u_i$ knows $I_D$.
\medskip

\subsection{Proof of Theorem~\ref{thm:lic_spanner}} \label{sec:proof_lic_spanner}

\paragraph{High-level Overview}
We start our analysis by deriving a lower bound on $\lic_\cG(\cA)$ in terms of the sum over the mutual information terms $\II[ G : \Pi_i \mid Y_i, Z_i, I_{D,Y_i}\!=\! 1]$, where the sum ranges over all $u_i \in U$.
In fact, since the critical edges incident to $u_i$ are fully determined by $G$, we can instead use
$\II[ C_i : \Pi_i \mid Y_i, Z_i, I_{D,Y_i}\!=\! 1]$, which tells us that the local information cost is bounded from below by the sum over the expected amount of information that each $u_i$'s transcript $\Pi_i$ reveals about its critical incident edges $C_i$, conditioned on $u_i$'s initial knowledge.
By definition, the mutual information is equivalent to how much the entropy of $C_i$ decreases upon revealing $\Pi_i$, which corresponds to the difference
\[
  \HH[ C_i \mid Y_i, Z_i, I_{D,Y_i}\!=\! 1] - \HH[ C_i \mid \Pi_i , Y_i, Z_i, I_{D,Y_i}\!=\! 1].
\]
Intuitively speaking, this is the difference between the \emph{initial entropy} of $C_i$, i.e., before executing the algorithm, and the \emph{remaining entropy} after $u_i$ has received all messages.
We can obtain a bound on the initial entropy that depends on the expected number of critical edges incident to $u_i$, see Lemma~\ref{lem:initial}.
As we explain in more detail below, we need a technical result (Lemma~\ref{lem:ISi}) to lift the bound into the probability space where we also condition on $I_{S_i}\!=\! 1$, as we need to compute the \emph{expected} difference between the initial and remaining entropy terms.
We can only hope to obtain a reasonably small upper bound on the remaining entropy if we assume that $u_i$ outputs few incident spanner edges, which motivates the conditioning on event $I_{S_i}\!=\! 1$, see Lemma~\ref{lem:final}.
Finally, we combine the bounds of Lemmas~\ref{lem:initial} and ~\ref{lem:final} to obtain the sought bound on $\lic_\cG(\cA)$.

\medskip

The proof of the following Lemma~\ref{lem:sparse} follows by a counting argument: if more than $n/\log n$ nodes had a degree (in the spanner) that is greater than $n^{\frac{2}{k+1}+ \frac{1}{8k^2} }$, we would obtain a contradiction to the assumed bound on the number of edges.
\newcommand{\lemSparse}{
Recall that $\sparse \subseteq U$ is the subset of nodes such that each $u \in \sparse$ outputs at most $n^{\frac{2}{k+1}+ \frac{1}{8k^2} }$ edges. If the algorithm terminates correctly, then
\begin{align}
  |\sparse| \ge |U| - \frac{n}{\log n} = \frac{n}{2} - \frac{n}{\log n}. %
  \notag
\end{align}

}
\begin{lemma} \label{lem:sparse}
  \lemSparse
\end{lemma}

\begin{proof}
  Assume towards a contradiction that $|U \setminus \sparse| \ge  \frac{ n}{\log n}$. %
  This means that a set $B$ of at least $\frac{n}{\log n}$ nodes in $U$ output at least $n^{\frac{2}{k+1}+\frac{1}{8k^2}}$ spanner edges each. %
  Consequently, the algorithm outputs at least
  \begin{align}
    |B|n^{\frac{2}{k+1}+\frac{1}{8k^2}}
    = \Omega\lb(n^{1 + \frac{2}{k+1}+\frac{1}{8k^2} - \frac{\log\log n}{\log n} }\rb) \label{eq:sparse1}
  \end{align}
  edges in total.
  From \eqref{eq:k}, we know that $k = O \lb( \lb( \frac{\log n}{\log\log n} \rb)^{1/3}\rb)$ and hence $k \le \sqrt{\frac{\log n}{24\log \log n}}$, for sufficiently large $n$, which implies that
  \[
  \frac{1}{8k^2} - \frac{\log\log n}{\log n} \ge \frac{1}{12k^2}.
  \]
  It follows that the right-hand side of \eqref{eq:sparse1} is at least
  \[
  \Omega\lb(n^{1 + \frac{2}{k+1}+\frac{1}{12k^2}}\rb) = \Omega\lb(n^{1 + \frac{1}{t}+\frac{1}{12(2t-1)^2}}\rb),
  \] 
  thus exceeding the assumed bound on the size of the spanner stipulated by Theorem~\ref{thm:lic_spanner} and resulting in a contradiction.
\end{proof}

Next, we prove a lower bound on the number of critical edges by leveraging Lemma~\ref{lem:reachability} and the upper bound on the incident red edges stated in Lemma~\ref{lem:dense}.

\newcommand{\lemConcentrationCritical}{
Recall that $I_{Y_i} \!=\! 1$ if and only if $u_i$ has at least $\tfrac{1}{16}n^{\frac{1}{k+1}}$ incident critical edges.
It holds that
\[
  \Pr[ I_{Y_i} \!=\! 1\mid I_{D} \!=\! 1]
  \ge
  1 - O \lb( n^{-\frac{1}{k^3}}\rb).
\]
}
\begin{lemma} \label{lem:concentration_critical}
  \lemConcentrationCritical
\end{lemma}

\begin{proof}
  Let ${Y_i}$ be the number of critical edges incident to $u_i$.
  We first give some intuition why the lemma holds.
Recall from Lemma~\ref{lem:reachability} that
$
  \EE[{Y_i}\mid I_D \!=\! 1] \ge \tfrac{3}{4}
  \lb(n^{\frac{1}{k+1}}
    - 2n^{\frac{1}{k+1} - \frac{1}{k^3}}
  \rb)
    - o(1).
    \notag
$
We know that ${Y_i}$ cannot exceed the number of red edges incident to $u_i$ and, conditioned on $I_D \!=\! 1$, Lemma~\ref{lem:dense} tells us that we can set the upper bound on $Y_i$ to be
$B = \tfrac{3}{4}n^{\frac{1}{k+1}}
+ c_1    n^{\frac{1}{2(k+1)}}\sqrt{\log n},$
for a suitable constant $c_1\ge 1$.
Consider the random variable $B - Y_i$, which is always positive due to the conditioning on  $I_D \!=\! 1$.
Applying Markov's inequality (c.f.\ Theorem~3.1 in \cite{upfalmitzenmacher}) to $B-Y_i$ (and hence in ``reverse'' direction to $Y_i$) shows that $Y_i$ is likely to be sufficiently large which is equivalent to event $I_{Y_i}\!=\! 1$.

	We now give the formal proof.
  From Lemma~\ref{lem:reachability}, we know that
  \begin{align}
    \EE[{Y_i}\mid I_D \!=\! 1] \ge \tfrac{3}{4}
    \lb(n^{\frac{1}{k+1}}
      - 2n^{\frac{1}{k+1} - \frac{1}{k^3}}
    \rb)
      - o(1).
      \label{eq:revMarkovEE}
  \end{align}

  We know that ${Y_i}$ cannot exceed the number of red edges incident to $u_i$ and, conditioned on $I_D \!=\! 1$ (i.e.\ event $\dense$ happens), Lemma~\ref{lem:dense} tells us that we can set the upper bound to be
  \begin{align}
  B = \tfrac{3}{4}n^{\frac{1}{k+1}}
  + c_1    n^{\frac{1}{2(k+1)}}\sqrt{\log n},
  \label{eq:revMarkovB}
  \end{align}
  for a suitable constant $c_1\ge 1$.
  Consider the random variable $B - Y_i$, which is always positive due to the conditioning on  $I_D \!=\! 1$.
  By Markov's inequality (c.f.\ Theorem~3.1 in \cite{upfalmitzenmacher}), it follows that, for any $a<B$,
  \[
    \Pr\lb[ Y_i \le a \ \middle|\ I_D \!=\! 1\rb]
    = \Pr\lb[ B-Y_i \ge B-a \ \middle|\ I_D \!=\! 1\rb]
    \le \frac{B - \EE[ Y_i \mid I_D \!=\! 1]}{B - a}.
  \]
  Choosing $a = \tfrac{1}{16}n^{\frac{1}{k+1}}$, and plugging \eqref{eq:revMarkovEE} and \eqref{eq:revMarkovB} into this concentration bound yields
  \begin{align}
    \Pr\lb[ {Y_i} \!\le\! \tfrac{1}{16}n^{\frac{1}{k+1}}
          \ \middle|\ I_D \!=\! 1\rb]
     &\le
       \frac{\lb(\tfrac{3}{4}n^{\frac{1}{k+1}}
       + c_1 n^{\frac{1}{2(k+1)} }\sqrt{\log n}\rb)
       -
       \tfrac{3}{4}
       \lb(n^{\frac{1}{k+1}}
         - 2n^{\frac{1}{k+1} - \frac{1}{k^3}}
       \rb)
         + o(1)}
       {\lb(\tfrac{3}{4}n^{\frac{1}{k+1}}
       + c_1 n^{\frac{1}{2(k+1)} }\sqrt{\log n}\rb) - \tfrac{1}{16}n^{\frac{1}{k+1}}}
       \notag\\
     &\le
     \frac{
     c_1n^{\frac{1}{2(k+1)} }\sqrt{\log n}
     +
     \tfrac{3}{2} n^{\frac{1}{k+1} - \frac{1}{k^3}}
       + o(1)}
     {\tfrac{11}{16}n^{\frac{1}{k+1}}
     + c_1 n^{\frac{1}{2(k+1)} }\sqrt{\log n}.
     }
     \notag\\
  \intertext{
    Observe that
      $n^{\frac{1}{2(k+1)} }\sqrt{\log n}
      =n^{\frac{1}{2(k+1)} + \frac{\sqrt{\log\log n}}{\log n}}
      \le n^{\frac{1}{2(k+1)} + \frac{1}{2k^2}}$,
      for any $k\ge 3$.
    Moreover,
      $n^{\frac{1}{2(k+1)} + \frac{1}{2k^2}}
      \le
       n^{\frac{1}{k+1} - \frac{1}{k^3}},
      $
    and therefore
  }
  \Pr\lb[ {Y_i} \!\le\! \tfrac{1}{16}n^{\frac{1}{k+1}}
          \ \middle|\ I_D \!=\! 1\rb]
    &\le
    \frac{
    \lt( c_1 + \frac{3}{2}  \rt) n^{\frac{1}{k+1} - \frac{1}{k^3}}
      + o(1)}
    {\tfrac{11}{16}n^{\frac{1}{k+1}}
    + c_1 n^{\frac{1}{2(k+1)} }\sqrt{\log n}
    }
    \notag\\
     &\le
     \frac{
     \lt( c_1 + \frac{3}{2}  \rt) n^{\frac{1}{k+1} - \frac{1}{k^3}}
       + o(1)}
     {\tfrac{11}{16}n^{\frac{1}{k+1}}
     }
     \notag\\
     &=
     O\lb(
        n^{-\frac{1}{k^3}}.
      \rb)
      \notag
  \end{align}
\end{proof}

We make use of the following conditioning property of mutual information:
\newcommand{\lemMutualConditioning}{
Let $X$, ${Y}$, $Z_1$, and $Z_2$ be discrete random variables.
Then, for any $z$ in the support of $Z_1$, it holds that
\[
  \II \lb[ X : {Y} \mid Z_1, Z_2 \rb]
    \ge \Pr[ Z_1 \!=\! z] \II \lb[ X : {Y} \mid Z_1 \!=\! z, Z_2 \rb].
\]
}
\begin{lemma} \label{lem:mutual_conditioning}
  \lemMutualConditioning
\end{lemma}

\begin{proof}
  \begin{align*}
    \II \lb[ X : {Y} \mid Z_1, Z_2 \rb]
      &= \sum_{z_1} \sum_{z_2}
          \Pr[ z_1, z_2 ]\ \II \lb[ X : {Y} \mid Z_1 \!=\! z_1, Z_2 \!=\! z_2 \rb] \\
      &= \sum_{z_1} \Pr[ z_1 ] \sum_{z_2}
          \Pr[ z_2 \mid z_1 ]\ \II \lb[ X : {Y} \mid Z_1 \!=\! z_1, Z_2 \!=\! z_2 \rb] \\
      &\ge \Pr[ z ]
          \II[ X : {Y} \mid Z_1 \!=\! z, Z_2 ],
  \end{align*}
  for any $z$ in the support of $Z_1$.
\end{proof}

 We are now ready to bound the local information cost $\lic_{\text{$\cG$}}(\cA)$ as stated in Definition~\ref{def:lic} (Page~\pageref{eq:lic_alg}). %
Since $U \subset V(G)$ and $X_i$ is just a shorthand for the random variables ${Y_i},Z_i,I_D,I_{Y_i}$, we have
\begin{align}
  \lic_{\text{$\cG$}}(\cA)
  &\ge
  \sum_{u_i \in U}^{} \II\lb[ G : \Pi_i \mid X_i\rb]
    \notag\\
  &=
  \sum_{u_i \in U}^{} \II\lb[ G : \Pi_i \mid {Y_i},Z_i,I_D,I_{Y_i} \rb]
    \notag\\
  \ann{by Lemma~\ref{lem:mutual_conditioning}} 
  &\ge
  \Pr[ I_D \!=\! 1 ]
  \sum_{u_i \in U}^{}
    \II\lb[ G : \Pi_i\ \middle|\ {Y_i},Z_i,I_{Y_i},I_D \!=\! 1  \rb]
    \notag\\
  &\ge
  \Pr[ I_D \!=\! 1 ]
  \sum_{u_i \in U}^{}
    \Pr[ I_{Y_i} \!=\! 1 \mid I_{D} \!=\! 1 ]
    \II\lb[ G : \Pi_i\ \middle|\ {Y_i},Z_i,I_{{D,{Y_i}}} \!=\! 1 \rb],
    \label{eq:lic_spanner_bnd10}
\end{align}
where, in the last step, we have applied Lemma~\ref{lem:mutual_conditioning} to each term of the sum.
Without conditioning on the initial states of the nodes, it holds that, for all $u_i,u_j \in U$,
\[
\Pr[ I_{Y_i} \!=\! 1 \mid I_{D} \!=\! 1 ] = \Pr[ I_{Y_j} \!=\! 1 \mid I_{D} \!=\! 1 ] \ge 1 - O \lb( n^{-\frac{1}{k^3}}\rb), %
\]
where the inequality follows from Lemma~\ref{lem:concentration_critical}.
Furthermore, we know from Lemma~\ref{lem:dense} that $\Pr[ I_D \!=\! 1 ] \ge 1 - \tfrac{1}{n}$ and hence
\begin{align*}
  \Pr[ I_D \!=\! 1 ] \Pr[ I_{Y_i} \!=\! 1 \mid I_{D} \!=\! 1 ] \ge \lb(1 - O \lb( n^{-\frac{1}{k^3}}\rb)\rb) \lb( 1 - \tfrac{1}{n}\rb) \ge \tfrac{1}{2},
\end{align*}
for sufficiently large $n$,
since we assume that $k = O \lb( \lb( \log n /\log\log n \rb)^{1/3} \rb)$.
Applying these observations to \eqref{eq:lic_spanner_bnd10}, we get
\begin{align}
\lic_{\text{$\cG$}}(\cA)
  &\ge
  \frac{1}{2}
  \sum_{u_i \in U }^{}
    \II\lb[ G : \Pi_i\ \middle|\ {Y_i},Z_i,I_{{D,{Y_i}}} \!=\! 1 \rb].
    \notag
\end{align}
  For each $u_i \in U$, its incident critical edges given by the random variable $\critical_i$, are determined by the graph $G$.
  This means that $\Pi_i \ra G \ra \critical_i$ forms a Markov chain and
  hence the data processing inequality (see Lemma~\ref{lem:dataprocessing}) implies that \[\II\lb[ G : \Pi_i\ \middle|\ {Y_i},Z_i,I_{{D,{Y_i}}} \!=\! 1 \rb] \ge \II\lb[ \critical_i : \Pi_i\ \middle|\ {Y_i},Z_i,I_{{D,{Y_i}}} \!=\! 1 \rb].\]
  Moreover, we can write the mutual information between $\critical_i$ and $\Pi_i$ in terms of the conditional entropies (see \eqref{eq:mutual2} in Appendix~\ref{sec:tools}), yielding
\begin{align}
\lic_{\text{$\cG$}}(\cA)
  &\ge
  \frac{1}{2}
  \sum_{u_i \in U }^{}
    \lb(
    \HH\lb[ \critical_i\ \middle|\ Y_i, Z_i, I_{{D,{Y_i}}}\!=\! 1 \rb]
    -
    \HH \lb[ \critical_i\ \middle|\ \Pi_i,
    {Y_i},Z_i,I_{{D,{Y_i}}} \!=\! 1 \rb]
    \rb).
    \label{eq:mutual2entropy}
  \end{align}

We will proceed by analyzing the entropy terms on the right-hand side in \eqref{eq:mutual2entropy}.
To bound the remaining entropy $\HH \lb[ \critical_i\ \middle|\ \Pi_i,
{Y_i},Z_i,I_{{D,{Y_i}}} \!=\! 1 \rb]$, i.e., after $u_i$ has received all messages, we need to reason about the number of spanner edges output by $u_i$.
We provide additional knowledge to the algorithm ``for free'', by revealing to each $u_i$ whether $u_i \in \sparse$ and $|\sparse| \ge \frac{n}{2} - \frac{n}{\log n}$, which is captured by the indicator random variable $I_{S_i}$ defined above.
Note that this assumption can only help the algorithm and hence does not weaken Theorem~\ref{thm:lic_spanner}.

However, the conditioning on $I_{S_i}$ introduces a technical challenge: We would like to compute the \emph{expected} difference between the two entropy terms, but we are computing the initial entropy of the critical edges (i.e. $\HH \lb[ \critical_i\ \middle|\ Y_i, Z_i, I_{{D,Y_i}} \!=\! 1 \rb]$) conditioned on $I_{{D,Y_i}} \!=\! 1$, whereas, for the remaining entropy, we also condition on $I_{S_i}\!=\! 1$.
Thus, we need the following lemma that will help us to lift the bound on the initial entropy to the probability space where we condition on all three indicator random variables.

\begin{lemma} \label{lem:ISi}
  $
  \Pr\lb[ I_{S_i} \!=\! 1 \middle|\ I_{{D,Y_i}} \!=\! 1\rb]
  \ge
   1 - O\lb(\frac{1}{\log n}\rb).
  $
\end{lemma}
\begin{proof}
  We begin our analysis by first deriving a lower bound on $\Pr\lb[ I_{S_i}\!=\! 1 \rb]$.
  Without conditioning on the initial states, $\Pr\lb[ u_i \!\in\!\sparse \rb] = \Pr\lb[ u_j \!\in\!\sparse \rb]$, for all $u_i, u_j \in U$, and hence it follows that
  \[
    \Pr\lb[ u_i \!\in\!\sparse\ \middle|\ |\sparse| \ge \tfrac{n}{2} - \tfrac{n}{\log n} \rb]
    \ge \frac{\tfrac{n}{2} - \tfrac{n}{\log n}}{\tfrac{n}{2}}
    = 1 - O\lb( \frac{1}{\log n} \rb).
  \]
  By assumption, the algorithm succeeds with probability at least $1 - \tfrac{1}{n}$ and, if it does, then it also holds that $|\sparse| \ge \tfrac{n}{2} - \tfrac{n}{\log n}$ according to Lemma~\ref{lem:sparse}.
  This means that
  \begin{align}
    \Pr\lb[ I_{S_i} \!=\! 1 \rb]
    &=
     \Pr\lb[ u_i \!\in\! \sparse\ \middle|\ |\sparse| \ge \tfrac{n}{2} - \tfrac{n}{\log n} \rb]
      \Pr\lb[ |\sparse| \ge \tfrac{n}{2} - \tfrac{n}{\log n} \rb] \notag\\ 
    &\ge 1 - O\lb( \frac{1}{\log n} \rb) - \frac{1}{n} = 1 - O\lb( \frac{1}{\log n} \rb).
    \notag
  \end{align}

  We now return to bounding $\Pr\lb[ I_{S_i} \!=\! 1 \ \middle|\ I_{D,Y_i} \!=\! 1 \rb]$.
  We have
  \begin{align}
      1 - O\lb( \frac{1}{\log n} \rb)
      &\le
      \Pr\lb[ I_{S_i} \!=\! 1 \rb] \notag\\ 
    &=
      \begin{multlined}[t]
      \Pr\lb[ I_{S_i} \!=\! 1 \ \middle|\ I_{D,Y_i} \!=\! 1 \rb]
      \Pr\lb[ I_{D,Y_i} \!=\! 1  \rb] \\
      +
      \Pr\lb[ I_{S_i} \!=\! 1 \ \middle|\ \neg I_{D,Y_i} \!=\! 1 \rb]
      \Pr\lb[ \neg I_{D,Y_i} \!=\! 1  \rb]
      \end{multlined}
    \notag \\
    &\le
      \Pr\lb[ I_{S_i} \!=\! 1 \ \middle|\ I_{D,Y_i} \!=\! 1 \rb]
      +
      \Pr\lb[ \neg I_{D,Y_i} \!=\! 1  \rb],
    \notag
  \end{align}
  and hence
  \[
    \Pr\lb[ I_{S_i} \!=\! 1 \ \middle|\ I_{D,Y_i} \!=\! 1 \rb]
    \ge
    1 - \Pr\lb[ \neg I_{D,Y_i} \!=\! 1  \rb]
      - O\lb( \frac{1}{\log n} \rb).
  \]
  To complete the proof of the lemma, we need to argue that $\Pr\lb[ \neg I_{D,Y_i} \!=\! 1  \rb] = O\lb( \frac{1}{\log n} \rb)$.
  We have
  \begin{align}
    \Pr\lb[ \neg I_{D,Y_i} \!=\! 1  \rb]
      &\le \Pr\lb[ I_D \!=\! 0 \rb]
       + \Pr\lb[ I_{Y_i} \!=\! 0 \ \middle|\ I_D \!=\! 1 \rb]
         \Pr\lb[ I_D \!=\! 1 \rb]\notag\\
      \ann{by Lemma~\ref{lem:dense}}
      &\le  \Pr\lb[ I_{Y_i} \!=\! 0 \ \middle|\ I_D \!=\! 1 \rb]
            + O\lb( \tfrac{1}{n} \rb)
        \notag\\
      \ann{by Lemma~\ref{lem:concentration_critical}}
      &= O\lb( \frac{1}{\log n} \rb).
        \notag
  \end{align}
\end{proof}

We are now ready to show a lower bound on the first entropy expression in \eqref{eq:mutual2entropy}. In more detail, we show that the initial entropy of the set of incident critical edges conditioned on the event $I_{D,Y_i}\!=\! 1$ is bounded from below by their expected number (ignoring logarithmic factors), where this expectation is conditional on the event that $u_i \in \sparse$ and $|\sparse| \ge \tfrac{n}{2} - \tfrac{n}{\log n}$ (i.e. $I_{S_i}\!=\! 1$) in addition to event $I_{D,Y_i} \!=\! 1$:

\begin{lemma} \label{lem:initial}
  Let $\critical_i$ denote the critical edges incident to $u_i$.
  It holds that
  \[
  \HH \lb[ \critical_i\ \middle|\ Y_i, Z_i, I_{{D,Y_i}} \!=\! 1 \rb]
    \ge
    \begin{multlined}[t]
    \lb(\frac{2}{k+1} + \frac{1}{4k^2} \rb) \lb(\log_2(n) - O(1)\rb)\cdot
    \EE \lb[ Y_i\ \middle|\ I_{{D,S_i,Y_i}} \!=\! 1 \rb] \\
    -
    \EE \lb[ Y_i \log_2 Y_i\ \middle|\ I_{{D,S_i,Y_i}} \!=\! 1 \rb].
    \end{multlined}
  \]
\end{lemma}
\begin{proof}
  By the definition of conditional entropy, we have
  \begin{align}
    \HH \lb[ \critical_i\ \middle|\ Y_i, Z_i, I_{{D,Y_i}} \!=\! 1 \rb] = \sum_{y,z}
      \Pr[ y,z \mid I_{{D,Y_i}} \!=\! 1 ]
       \HH \lb[ \critical_i\ \middle|\ Y_i \!=\! y, Z_i \!=\! z, I_{{D,Y_i}} \!=\! 1 \rb],
       \label{eq:initial1}
  \end{align}
  where $y$ and $z$ are such that $\Pr[Y_i \!=\! y,Z_i \!=\! z \mid I_{D,Y_i}\!=\! 1] > 0$.

  We will derive a bound on $\HH \lb[ \critical_i\ \middle|\ Y_i \!=\! y, Z_i \!=\! z, I_{{D,Y_i}} \!=\! 1 \rb]$.
  Recall from Section~\ref{sec:lbgraph} that the assignment of the random IDs to the nodes is done independently of the sampling of the red edges (which determines $I_D$ and $I_{Y_i}$), and hence the assignment of the IDs to the nodes in the neighborhood of $u_i$ (given by $z$) is independent of $I_D$ and $I_{Y_i}$.
  In other words, the ID assignments given to $u_i$'s neighbors which is known to $u_i$ does not reveal any information about which edges are blue and which ones are red, let alone which edges are critical.
  More formally, for each neighbor $w \in U \cup V$ of $u_i$, the edge $(u_i,w)$ is red with some probability $p$ (independent of $w$'s ID), and critical with some probability $p' \le p$, where this event is also independent of $w$'s ID.
  Therefore, if we consider any two subsets $c$ and $c'$ of exactly $y$ nodes chosen from the neighborhood $z$ of $u_i$ that are identified by their (unique) IDs, it follows from the above that
  $
    \Pr[ \critical_i \!=\! c \mid y,z,I_{D,Y_i}\!=\! 1]
    =
    \Pr[ \critical_i \!=\! c' \mid y,z,I_{D,Y_i}\!=\! 1].
    \label{eq:prob1}
  $

  To obtain a lower bound on $u_i$'s degree, we recall that $u_i$ has $n^{\frac{2}{k+1} + \frac{1}{4k^2}}$ incident (blue) edges, and hence
  \begin{align}
    \Pr[ \critical_i \!=\! c \mid y, z, I_{{D,{Y_i}}} \!=\! 1 ] \le 1 / {n^{\frac{2}{k+1} + \frac{1}{4k^2}} \choose y}. \label{eq:prob2}
  \end{align}
  We combine these observations to obtain that
  \begin{align}
    \HH \lb[ \critical_i \mid y, z, I_{{D,{Y_i}}} \!=\! 1 \rb]
      &=
         \sum_c \Pr[ \critical_i \!=\! c \mid y, z, I_{{D,{Y_i}}} \!=\! 1 ] \log_2 \lb( 1/\Pr[ c \mid y, z, I_{{D,{Y_i}}} \!=\! 1 ] \rb) \notag\\
\ann{by  \eqref{eq:prob2}}
      &\ge
       \sum_c \Pr[ c \mid y, z, I_{{D,{Y_i}}} \!=\! 1 ] \log_2 {n^{\frac{2}{k+1} + \frac{1}{4k^2}} \choose y} \notag\\
      &=
      \log_2 {n^{\frac{2}{k+1} + \frac{1}{4k^2}} \choose y}
        \sum_c \Pr[ c \mid y, z, I_{{D,{Y_i}}} \!=\! 1 ]
             \notag\\
      &=
      \log_2 {n^{\frac{2}{k+1} + \frac{1}{4k^2}} \choose y}
          \notag
          \\
\ann{\text{since ${n \choose k} \ge \lb( \frac{n}{k} \rb)^k$}}
      &\ge
      \log_2 \lb( \frac{n^{\frac{2}{k+1} + \frac{1}{4k^2}}}{y} \rb)^{y}
         \notag 
                  \\
      &= \lb( \frac{2}{k+1} + \frac{1}{4k^2} \rb) y \log_2 n - y\log_2 y.\notag
  \end{align}
  Applying this bound to \eqref{eq:initial1}, we get
  \begin{align}
    \HH \lb[ \critical_i\ \middle|\ Y_i, Z_i, I_{{D,Y_i}} \!=\! 1 \rb] &\ge \sum_{y,z}
      \Pr[ y,z \mid I_{{D,Y_i}} \!=\! 1 ]
       \lb(\lb( \frac{2}{k+1} + \frac{1}{4k^2} \rb) y \log_2 n - y\log_2 y\rb)
     \notag\\
    &= \sum_{y}
      \Pr[ y \mid I_{{D,Y_i}} \!=\! 1 ]
       \lb(\lb( \frac{2}{k+1} + \frac{1}{4k^2} \rb) y \log_2 n - y\log_2 y\rb).
       \label{eq:initial2}
  \end{align}
  The final expression is equivalent to
  $\EE_{Y_i} \lb[ \lb( \frac{2}{k+1} + \frac{1}{4k^2} \rb) Y_i \log_2 n - Y_i\log_2 Y_i \ \middle|\ I_{{D,Y_i}} \!=\! 1 \rb]$.

  As explained earlier, we need to compute the expectation conditioned on $I_{{D,S_i,Y_i}} \!=\! 1$.
  We will use Lemma~\ref{lem:ISi} to lift \eqref{eq:initial2} to this probability space.
  For any $y$ in the support of $Y_i$, we have
  \begin{align}
    \Pr\lb[ y\ \middle|\ I_{{D,Y_i}} \!=\! 1\rb]
    &\ge
    \Pr\lb[ y\ \middle|\ I_{{D,S_i,Y_i}} \!=\! 1\rb]
    \Pr\lb[ I_{S_i} \!=\! 1 \middle|\ I_{{D,Y_i}} \!=\! 1 \rb]
    \notag\\
\ann{by Lem.~\ref{lem:ISi}}
    &\ge
    \lb( 1 - O\lb(\frac{1}{\log n}\rb) \rb)
    \Pr\lb[ y\ \middle|\ I_{{D,S_i,Y_i}} \!=\! 1\rb].
    \notag
  \end{align}
  Returning to \eqref{eq:initial2}, we conclude that
  \begin{align}
    \HH [ \critical_i &\mid Y_i, Z_i, I_{{D,{Y_i}}} \!=\! 1 ] \notag\\
    &\ge
      \lb( 1 - O\lb(\frac{1}{\log n}\rb) \rb)
      \sum_{y}
        \Pr[ y \mid I_{{D,S_i,Y_i}} \!=\! 1 ]
         \lb(\lb( \frac{2}{k+1} + \frac{1}{4k^2} \rb) y \log_2 n - y\log_2 y\rb)\notag\\
    &=
      \lb( 1 - O\lb(\frac{1}{\log n}\rb) \rb)\cdot
      \EE_{Y_i} \lb[
        \lb( \frac{2}{k+1} + \frac{1}{4k^2} \rb) Y_i \log_2 n - Y_i\log_2 Y_i
        \ \middle|\
        I_{{D,S_i,Y_i}} \!=\! 1
      \rb].\notag
  \end{align}
  The result follows by linearity of expectation.
\end{proof}

In the next lemma, we will derive an upper bound on the second entropy expression in \eqref{eq:mutual2entropy}.
In particular, we show that the entropy that remains after $u_i$ has received all messages that were sent to it by the algorithm is sufficiently small in the case where $u_i$ has few incident spanner edges, i.e., $I_{S_i}\!=\! 1$.

\begin{lemma} \label{lem:final}
  Recall that $\critical_i$ denotes the critical edges incident to $u_i$.
  We have
  \begin{multline}
  \HH \lb[ \critical_i\ \middle|\ \Pi_i,Y_i,Z_i,I_{{D,Y_i}} \!=\! 1 \rb]
    \le
    \lb(\lb( \frac{2}{k+1} + \frac{1}{8k^2} \rb)  \log_2 n + \log_2 e\rb)
    \EE\lb[
      Y_i
     \ \middle|\ I_{{D,S_i,Y_i}} \!=\! 1 \rb]
     \\
     +
     O\lb(\frac{3}{k+1}\rb)
     \EE\lb[
       {Y_i}
       \ \middle|\ I_{{D,Y_i}} \!=\! 1, I_{S_i} \!=\! 0 \rb]
     -
     \EE\lb[
      Y_i \log_2 \lb( {Y_i} \rb)
      \ \middle|\ I_{{D,S_i,Y_i}} \!=\! 1 \rb].
      \notag
  \end{multline}
\end{lemma}
\begin{proof}
  As outlined above, we can deduce $I_{S_i}$ from the transcript because the first bit that $u_i$ receives is the value of $I_{S_i}$.
  Moreover, the set of incident spanner edges $F_i$ that are finally output by $u_i$ is determined by the transcript $\Pi_i$ and the initial state of $u_i$ as given by $Y_i$, $Z_i$ (including $u_i$'s private random bits).
  Hence,
  \begin{align}
    \HH [ &\critical_i\ \mid\ \Pi_i,Y_i,Z_i,I_{{D,Y_i}} \!=\! 1 ] \notag\\ 
    &\le
      \HH \lb[ \critical_i\ \middle|\ F_i,I_{S_i},Y_i,Z_i,I_{{D,Y_i}} \!=\! 1 \rb] \notag\\
\ann{by Lem.~\ref{lem:entropy_conditioning}} 
    &\le
      \HH \lb[ \critical_i\ \middle|\ F_i,I_{S_i},Y_i,I_{{D,Y_i}} \!=\! 1 \rb] \notag\\
    &=
      \sum_{b \in \set{0,1}}
      \sum_{f,y}
      \Pr \lb[ F_i \!=\! f, I_{S_i}\!=\! b, Y_i \!=\! y\ \middle|\ I_{{D,Y_i}} \!=\! 1\rb]
      \HH \lb[ \critical_i\ \middle|\ f, b, y, I_{{D,Y_i}} \!=\! 1 \rb]
      \notag\\
    &=
      \begin{multlined}[t]
      \Pr \lb[ I_{S_i}\!=\! 1\ \middle|\ I_{{D,Y_i}} \!=\! 1\rb]
      \sum_{f,y}
      \Pr \lb[ f,y \ \middle|\ I_{{D,S_i,Y_i}} \!=\! 1 \rb]
      \HH \lb[ \critical_i\ \middle|\ f,y, I_{{D,S_i,Y_i}} \!=\! 1 \rb] \\
      +
      \Pr \lb[ I_{S_i}\!=\! 0\ \middle|\ I_{{D,Y_i}} \!=\! 1\rb]
      \sum_{f,y} \Big(
      \Pr \lb[ f,y \ \middle|\ I_{{D,Y_i}} \!=\! 1, I_{S_i} \!=\! 0 \rb] \\
      \cdot\HH \lb[ \critical_i\ \middle|\ f, y, I_{{D,Y_i}} \!=\! 1, I_{S_i} \!=\! 0 \rb] \Big)
      \end{multlined}
      \notag \\
    &\le
      \sum_{f,y}
      \Pr \lb[ f,y \ \middle|\ I_{{D,S_i,Y_i}} \!=\! 1 \rb]
      \HH \lb[ \critical_i\ \middle|\ f,y, I_{{D,S_i,Y_i}} \!=\! 1 \rb] \notag\\
      &\phantom{-}+
      O\lb( \frac{1}{\log n} \rb)
      \sum_{f,y}
            \Pr \lb[ f,y \ \middle|\ I_{{D,Y_i}} \!=\! 1, I_{S_i} \!=\! 0 \rb]
            \HH \lb[ \critical_i\ \middle|\ f,y, I_{{D,Y_i}} \!=\! 1, I_{S_i} \!=\! 0 \rb] \label{eq:final_two_sums}
  \end{align}
  where the last inequality holds because Lemma~\ref{lem:ISi} implies that $\Pr\lb[ I_{S_i}\!=\! 0\ \middle|\ I_{{D,Y_i}} \!=\! 1\rb] = O\lb( {1}/{\log n} \rb)$.

  We will separately bound the sums on the right-hand side.
  Fix any $f$ and $y$.
  To obtain an upper bound on $\HH \lb[ \critical_i\ \middle|\ f,y, I_{{D,S_i,Y_i}} \!=\! 1 \rb]$, recall that the conditioning on $I_{S_i} \!=\! 1$ says that $u_i$'s output $f$ contains at most $n^{\frac{2}{k+1} + \frac{1}{8k^2}}$ spanner edges (see Lemma~\ref{lem:sparse})
  and, we know from Lemma~\ref{lem:keepEdges} that every critical edge must be part of the spanner, which tells us that the $y$ critical edges incident to $u_i$ are part of $u_i$'s output.
  The total number of ways we can choose a subset $c$ of size $y$ from the edges in $f$ is
  \[
  {|f| \choose y} \le
  {n^{\frac{2}{k+1} + \frac{1}{8k^2}} \choose y}.
  \]
  Moreover, the (remaining) entropy $\HH \lb[ \critical_i\ \middle|\ f,y, I_{{D,S_i,Y_i}} \!=\! 1 \rb]$ is maximized if the conditional distribution of $\critical_i$ is uniform over all subsets $c$ of size $y$.
  Combining these observations, we get
  \begin{align}
    \HH \lb[ \critical_i\ \middle|\ f,y, I_{{D,S_i,Y_i}} \!=\! 1 \rb]
    &=
       \sum_c \Pr[ \critical_i \!=\! c \mid f,y, I_{{D,S_i,Y_i}} \!=\! 1] \log_2 \lb( \frac{1}{\Pr[ c \mid f,y,  I_{{D,S_i,Y_i}} \!=\! 1 ]} \rb) \notag\\
    &\le
\log_2 {n^{\frac{2}{k+1} + \frac{1}{8k^2}} \choose y} 
      \sum_c \Pr\lb[ \critical_i \!=\! c \mid f,y, I_{{D,S_i,Y_i}} \!=\! 1\rb] \notag\\
    &=
    \log_2 {n^{\frac{2}{k+1} + \frac{1}{8k^2}} \choose y} \notag
                \\
    \ann{\text{since ${n \choose k} \le \lb( \frac{en}{k} \rb)^k$}}
    &\le
    \log_2 \lb( \frac{e\ n^{\frac{2}{k+1} + \frac{1}{8k^2}}}{y} \rb)^{y}
                 \notag\\
    &= y \lb(\lb( \frac{2}{k+1} + \frac{1}{8k^2} \rb)  \log_2 n +  \log_2 \lb( \frac{e}{y} \rb)\rb).\label{eq:final_first_entropy}
  \end{align}

  Next, we derive an upper bound on the entropy term in the second sum of \eqref{eq:final_two_sums}, i.e., $\HH \lb[ \critical_i\ \middle|\ f,y, I_{{D,Y_i}} \!=\! 1, I_{S_i} \!=\! 0 \rb]$.
  Since $I_{S_i} \!=\! 0$, we cannot make any assumptions on the number of edges that are included in the output of $u_i$;
  in fact, it may happen that $u_i$ simply outputs all incident edges and so the entropy of $\critical_i$ may not decrease at all.
  Since we condition on $I_D = 1$, we know from Lemma~\ref{lem:dense} that the maximum number of blue and red edges incident to $u_i$ is bounded by
  \begin{align}
    d_\text{max} = n^{\frac{2}{k+1} + \frac{1}{4k^2}} + n^{\frac{1}{k+1}}
                 \le n^{\frac{3}{k+1}}, %
                 \label{eq:dmax}
  \end{align}
  for sufficiently large $n$.
  The conditioning on $Y_i \!=\! y$ ensures that $u_i$ has exactly $y$ critical edges, which means that the entropy is maximized if the distribution of $\critical_i$ over the at most $d_\text{max}$ edges incident to $u_i$ is uniform with probability $1/{d_\text{max} \choose y}$.
  That is,
  \begin{align}
    \HH[ &\critical_i\ \mid f,y, I_{{D,Y_i}} \!=\! 1, I_{S_i} \!=\! 0 ] \\
    &= \sum_{c}^{}
       \Pr\lb[ c \ \middle|\ f,y,I_{{D,Y_i}} \!=\! 1, I_{S_i}\!=\! 0\rb]
       \log_2\lb( \frac{1}{\Pr\lb[ c \ \middle|\ f,y,I_{{D,Y_i}} \!=\! 1, I_{S_i}\!=\! 0\rb]}\rb) \notag\\
    &\le \sum_{c}^{}
       \Pr\lb[ c \ \middle|\ f,y,I_{{D,Y_i}} \!=\! 1, I_{S_i}\!=\! 0\rb]
       \log_2 {d_{\text{max}} \choose y} \notag\\
\ann{since ${N \choose K} \le N^K$}
    &\le
       {y}\log_2 d_\text{max}
          \sum_c \Pr\lb[ c \ \middle|\ f,y,I_{{D,Y_i}} \!=\! 1, I_{S_i}\!=\! 0\rb]
          \notag\\
    &\le
       \frac{3y}{k+1}\log_2 n,
       \label{eq:final_second_entropy}
  \end{align}
  where the last step follows from \eqref{eq:dmax} and the fact that $\sum_c \Pr\lb[ c \mid \dots\rb] = 1$.
  \enlargethispage{2\baselineskip}

  We now combine the bounds that we obtained in \eqref{eq:final_first_entropy} and \eqref{eq:final_second_entropy} to obtain an upper bound on the expression $\HH \lb[ \critical_i\ \middle|\ \Pi_i,Y_i,Z_i,I_{{D,Y_i}} \!=\! 1 \rb]$.
  From \eqref{eq:final_two_sums}, it follows that
  \begin{align}
    \HH [ \critical_i &\mid \Pi_i,Y_i,Z_i,I_{{D,Y_i}} \!=\! 1 ] \notag\\
    &\le
      \begin{multlined}[t]
      \sum_{f,y}
      \Pr \lb[ f,y \ \middle|\ I_{{D,S_i,Y_i}} \!=\! 1 \rb]
      y \lb(\lb( \frac{2}{k+1} + \frac{1}{8k^2} \rb)  \log_2 n +  \log_2 \lb( \frac{e}{y} \rb)\rb) \\
      +
      O\lb( \frac{1}{\log n}\rb)
      \sum_{f,y}
            \Pr \lb[ f,y \ \middle|\ I_{{D,Y_i}} \!=\! 1, I_{S_i} \!=\! 0 \rb]
      \frac{3y}{k+1}\log_2 n 
      \end{multlined} \notag\\
    &=
      \EE_{F_i,Y_i}\lb[
        Y_i
        \lb(\lb( \frac{2}{k+1} + \frac{1}{8k^2} \rb)  \log_2 n +  \log_2 \lb( \frac{e}{Y_i} \rb)\rb)
       \ \middle|\ I_{{D,S_i,Y_i}} \!=\! 1 \rb]
      \notag
      \\
      &\phantom{-----}+
      O\lb( \frac{1}{\log n}\rb)
      \EE_{F_i,Y_i}\lb[
        \frac{3Y_i}{k+1}\log_2 n
        \ \middle|\ I_{{D,Y_i}} \!=\! 1, I_{S_i} \!=\! 0
      \rb]\notag\\
    &=
      \lb(\lb( \frac{2}{k+1} + \frac{1}{8k^2} \rb)  \log_2 n  + \log_2 e\rb)
      \EE_{Y_i}\lb[
        Y_i
       \ \middle|\ I_{{D,S_i,Y_i}} \!=\! 1 \rb]
       \notag\\
       &\phantom{--}
      +
      O\lb(\frac{3}{k+1} \rb)
      \EE_{Y_i}\lb[
        {Y_i}
        \ \middle|\ I_{{D,Y_i}} \!=\! 1, I_{S_i} \!=\! 0
      \rb]
      -
      \EE_{Y_i}\lb[
       Y_i \log_2 \lb( {Y_i} \rb)
       \ \middle|\ I_{{D,S_i,Y_i}} \!=\! 1 \rb].\notag
  \end{align}
\end{proof}

Equipped with Lemmas~\ref{lem:initial} and \ref{lem:final}, we can continue our derivation of a lower bound on $\lic_\cG(\cA)$.
By applying these bounds to the two entropy terms on the right-hand side of \eqref{eq:mutual2entropy}, we get
\begin{align}
  \lic_\cG(\cA)
    &\ge 
    \begin{multlined}[t]
    \frac{1}{2}
    \sum_{u_i \in U}
    \bigg(
    \Big(
      \lb(\frac{1}{4k^2} - \frac{1}{8k^2}\rb)\log_2 n \\
      - O\lb( \frac{2}{k+1} + \frac{1}{4k^2} \rb) - \log_2 e
    \Big)
      \EE \lb[ Y_i \ \middle|\ I_{{D,S_i,Y_i}}\!=\! 1 \rb] 
    \end{multlined}
      \notag\\
    &\phantom{----------------}
      -
      O\lb( \frac{3}{k+1}  \rb)
      \EE \lb[ Y_i\ \middle|\ I_{{D,Y_i}} \!=\! 1, I_{S_i} \!=\! 0 \rb]
     \bigg)\notag\\
    &\ge 
    \begin{multlined}[t]
    \frac{1}{2}
    \sum_{u_i \in U}
    \Big(
      \frac{1}{16k^2} \log_2 n\cdot
      \EE \lb[ Y_i \ \middle|\ I_{{D,S_i,Y_i}}\!=\! 1 \rb] \\
      -
      O\lb( \frac{3}{k+1}  \rb)
      \EE \lb[ Y_i\ \middle|\ I_{{D,Y_i}} \!=\! 1, I_{S_i} \!=\! 0 \rb]
     \Big),
    \end{multlined}
\notag
\end{align}
where we have used the fact that
\[
  \frac{1}{16k^2}\log_2 n \ge O\lb( \frac{2}{k+1} + \frac{1}{4k^2} \rb) +  \log_2(e),
\]
for sufficiently large $n$.
The conditioning on $I_{Y_i} \!=\! 1$ tells us that $Y_i \ge \tfrac{1}{16}n^{\frac{1}{k+1}}$ and hence also
\[
  \EE \lb[ Y_i \ \middle|\ I_{{D,S_i,Y_i}}\!=\! 1 \rb] \ge \tfrac{1}{16}n^{\frac{1}{k+1}}.
\]
On the other hand, the conditioning on $I_D \!=\! 1$ guarantees that the number of red edges incident to $u_i$ is at most $n^{\frac{1}{k+1}}$ (see Lemma~\ref{lem:dense}) and hence the same bound holds for the number of critical edges, which guarantees that
\[
  \EE \lb[ Y_i\ \middle|\ I_{{D,Y_i}} \!=\! 1, I_{S_i} \!=\! 0 \rb] \le n^{\frac{1}{k+1}}.
\]
From this we conclude that
\begin{align}
  \lic_\cG(\cA)
  &\ge \frac{|U|}{2}
  \lb(
    \frac{1}{16^2k^2} n^{\frac{1}{k+1}}\log_2 n
    - O\lb( n^{\frac{1}{k+1}} \rb)
  \rb) \notag\\
\ann{since $|U|=\Omega(n)$ and $k = O\lb( \lb(\frac{\log n}{\log\log n}\rb)^{1/3} \rb)$}
  &= \Omega \lb( \frac{1}{k^2} n^{1+\frac{1}{k+1}}\log n \rb)
  \notag \\
  &= \Omega \lb( \frac{1}{t^2} n^{1+\frac{1}{2t}}\log n \rb).
    \tag{by \eqref{eq:k}}
\end{align}
This completes the proof of Theorem~\ref{thm:lic_spanner}.

\subsection{Lower Bounds for Distributed Spanner Algorithms} \label{sec:applications}
In this section, we derive communication and time lower bounds from Theorem~\ref{thm:lic_spanner}.
We first apply Lemma~\ref{lem:conversion_async} to obtain the claimed bound on the communication complexity in the asynchronous model:

\begin{theorem} \label{thm:ASYNC}
Any algorithm that, with high probability, constructs a $(2t-1)$-spanner with $O\lb(n^{1+\frac{1}{t} + \frac{1}{16(2t-1)^2}}\rb)$ edges, for $2 \le t \le O\lb( \lb( {\log(n)}/{\log\log n} \rb)^{1/3} \rb)$, has a communication complexity of $\Omega\lb(\tfrac{1}{t^2} n^{1+\frac{1}{2t}} \log n\rb)$ bits in the asynchronous message passing clique under the $\ktone$ assumption.
\end{theorem}

From Lemma~\ref{lem:conversion_sync}, we get a similar result for the synchronous $\congest$ $\ktone$ model and the congested clique.
We do not explicitly state our bounds in terms of the message complexity, as this would change the result only by a logarithmic factor.

\begin{theorem} \label{thm:CONGESTKTONE}
  Consider the synchronous $\ktone$ congested clique and any
  $\tau = O\lb(\poly(n)\rb)$.
  Any $\tau$-round algorithm algorithm that, with high probability, outputs a $(2t-1)$-spanner with at most $O\lb(n^{1+\frac{1}{t} + \frac{1}{16(2t-1)^2}}\rb)$ edges sends at least $\Omega\lb(\tfrac{1}{t^2 \log n}\cdot n^{1+\frac{1}{2t}} \rb)$ bits in the worst case, for $2 \le t \le O\lb( \lb( {\log(n)}/{\log\log n} \rb)^{1/3} \rb)$.
  The same result holds in the $\congest$ $\ktone$ model.
\end{theorem}

When considering the $\congest$ $\ktone$ model and $t = O(1)$,  Theorem~\ref{thm:CONGESTKTONE} implies that, for any $(2t-1)$-spanner algorithm that succeeds with high probability and takes a polynomial number of rounds, there exists a graph where at least $\Omega\lb(n^{1+\epsilon}\rb)$ bits are sent, for some constant $\epsilon>0$.
On the other hand, we know from \cite{baswana} that the time complexity of $(2t-1)$-spanners is $O(t^2)$ rounds and, according to \cite{derbel}, $\Omega(t)$ is a lower bound even in the more powerful \textsf{LOCAL} model, which means that a time-optimal algorithm does not need to depend on $n$ at all!
Furthermore, by leveraging the time-encoding trick mentioned in Section~\ref{sec:intro} (and described in more detail in Appendix~\ref{sec:time-encoding}) it is possible to send only $\tilde O(n)$ bits at the cost of a larger running time, which matches the trivial lower bound on the communication complexity of $\Omega(n)$ bits  up to polylogarithmic factors.
Combining these observations we have the following:

\begin{corollary} \label{cor:simultaneous}
 There is no $(2t-1)$-spanner algorithm in the $\congest$ $\ktone$ model (or the congested clique) that outputs at most $O\lb(n^{1+\frac{1}{t} + \frac{1}{16(2t-1)^2}}\rb)$ edges with high probability and simultaneously achieves optimal time and optimal communication complexity.
\end{corollary}

Notice that Corollary~\ref{cor:simultaneous} reveals a gap between constructing a spanner and the problem of finding a minimum spanning tree (MST) in the congested clique, as the work of \cite{hegeman2015toward} shows that it is possible to solve the latter problem in $O(\polylog(n))$ time while sending only $\tilde O(n)$ bits.

We now turn our attention towards time complexity in the node-capacitated clique and the gossip model with limited bandwidth, as studied in \cite{DBLP:conf/podc/HaeuplerMS18}.
Applying Lemma~\ref{lem:conversion_nodecongested} reveals that constructing a spanner is harder than MST in the {node-capacitated clique}, for which there is a $O(\polylog(n))$ time algorithm (see \cite{DBLP:conf/spaa/AugustineGGHSKL19}):

\begin{theorem} \label{thm:NODECONGESTEDCLIQUE}
  Consider the node-capacitated clique defined in \cite{DBLP:conf/spaa/AugustineGGHSKL19}.
  Constructing a $(2t-1)$-spanner that, with high probability, has $O\lb(n^{1+\frac{1}{t} + \frac{1}{16(2t-1)^2}}\rb)$ edges requires $\Omega\lb(\tfrac{1}{t^2\log^4 n} n^{\frac{1}{2t}} \rb)$ rounds, for $2 \le t \le O\lb( \lb( {\log(n)}/{\log\log n} \rb)^{1/3} \rb)$.
\end{theorem}

From Lemma~\ref{lem:conversion_gossip} we get a similar result for the gossip model:

\begin{theorem} \label{thm:GOSSIP}
  Consider the push-pull gossip model where the link bandwidth is limited to $O(\log n)$ bits.
  Constructing a $(2t-1)$-spanner that, with high probability, has $O\lb(n^{1+\frac{1}{t} + \frac{1}{16(2t-1)^2}}\rb)$ edges requires $\Omega\lb(\tfrac{1}{t^2\log^3 n} n^{\frac{1}{2t}} \rb)$ rounds, for $2 \le t \le O\lb( \lb( {\log(n)}/{\log\log n} \rb)^{1/3} \rb)$.
\end{theorem}

\section{Future Work and Open Problems}
\label{sec:future}

We conclude by listing some interesting open problems and possible future research directions.

We have argued in Section~\ref{sec:applications} that it is impossible to construct a spanner in a way that is both time- and communication-optimal. However, this question is still unresolved for the basic problem of constructing minimum spanning trees:

\begin{open_problem} \label{open:MST}
  Is it possible to construct a minimum spanning tree in $\tilde O(D + \sqrt{n})$ rounds and with $O(n\polylog(n))$ communication complexity in the synchronous $\congest$ $\ktone$ model?
\end{open_problem}

We have shown in Section~\ref{sec:lic} that the local information cost $\lic_\gamma(P)$ presents a lower bound on the communication complexity $\CC_\gamma(P)$ of solving $P$ in the asynchronous message passing model with error at most $\gamma$.
A natural question is whether this relationship is tight (up to polylogarithmic factors) for all graph problems.

\begin{open_problem}
  Does it hold that $\CC_\gamma(P) = \tilde\Theta(\lic_\gamma(P))$, for every graph problem $P$?
\end{open_problem}

In this work, we have studied communication complexity lower bounds for multiplicative $(2t-1)$-spanners. To the best of our knowledge, all existing distributed spanner algorithms send at least $\Omega(t m)$ bits and hence it is still an open question whether the $\Omega(m)$ barrier known to hold for the clean network model (see Section~\ref{sec:intro}) also holds under the $\ktone$ assumption.
In fact, this question is still open even if we allow a polynomial number of rounds:

\begin{open_problem}
  Does there exist a distributed algorithm that constructs a $(2t-1)$-spanner with $O\lb(n^{1+\frac{1}{t}+\epsilon}\rb)$ edges, while sending $\tilde O\lb(n^{1+\frac{1}{2t}}\rb)$ bits and terminating in at most $O(\poly(n))$ rounds in the synchronous $\congest$ $\ktone$ model?
\end{open_problem}

\appendix

\section{A Simple Time-Encoding Mechanism for Obtaining Optimal Communication Complexity in the $\boldmath\congest$ $\ktone$ Model}
\label{sec:time-encoding}

We describe a folklore technique for obtaining a message-optimal algorithm for \emph{any} problem $P$ in the $\congest$ $\ktone$ model.
We first construct a spanning tree by using the algorithm of \cite{DBLP:conf/podc/KingKT15}, which takes time $\tilde O(n)$ and requires sending $\tilde O(n)$ bits.
Then, we elect a leader on this tree, which requires $\tilde O(n)$ bits using the algorithm of \cite{jacm15}.
Subsequently, the leader $u_*$ serves as the root of the spanning tree $T$ and every node knows its distance in $T$ from $u_*$.
Assuming an ID range of size $n^c$, there are at most $t=  2^{{n \choose 2}}(n!)\cdot{n^c \choose n}$ possible $n$-node graphs where a subset of $n$ IDs is chosen and assigned to the nodes by selecting one of the possible $n!$ permutations.
Let $E$ be some arbitrary enumeration of the set of permutations and let $E_i$ refer to the $i$-th item in the order stipulated by $E$.
We split the computation into iterations of $t$ rounds with the goal of performing a convergecast.
Let $d$ be the maximum distance of a node from the root $u_*$ in $T$, which can be known by all nodes in $O(D)$ additional rounds.

In the first iteration, each leaf at distance $d$ sends exactly $1$ bit to its parent in round $k$ if its local neighborhood corresponds to $E_k$.
Similarly, in iteration $i>1$, every node $u$ at distance $d-i+1$, sends $1$ bit in round $k'$ such that $E_{k'}$ corresponds to the subgraph consisting of $u$'s neighborhood as well as the topology information received from its children in the previous iterations.
Proceeding in this manner ensures that a node can convey its current knowledge of the network topology to its parent by sending only a single bit during one of the rounds in this iteration while remaining silent in all others. After $d$ iterations, the entire topological information is received by the root $u_*$, and this information again corresponds to some $E_{k''}$.
Subsequently we use another $d$ iterations, each consisting of $t$ rounds, to disseminate $E_{k''}$ starting from $u_*$ to all nodes in the network, and, similarly to the mechanism described above, this can be done by sending only $O(n)$ bits in total.
Finally, each node locally computes its output from $E_{k''}$ according to the solution to problem $P$.

\section{Tools from Information Theory.}
We first introduce some basic definitions from information theory such as entropy and mutual information and restate some facts (without proofs) that we use throughout the paper.
More details can be found in \cite{info_book}.

Throughout this section, we assume that $X$, $Y$, and $Z$ are discrete random variables.
We use capitals to denote random variables and corresponding lowercase characters for values.
To shorten the notation, we abbreviate the event ``$X\!=\! x$'' by simply writing ``$x$'', for random variables $X$ and a value $x$.%
\footnote{For instance, the expression $\HH \lb[ \critical_i \mid F,  y_i\rb]$ is the same as $\HH \lb[ \critical_i \mid F, {Y_i} \!=\! y_i\rb]$, where $C_i$, $F$, and $Y_i$ are random variables.}
When computing expected values, we sometimes use the subscript notation $\EE_X$ to clarify that the expectation is taken over the distribution of random variable $X$.

\begin{definition} \label{def:entropy}
  The \emph{entropy of $X$} is defined as
  \begin{align}
    \HH[ X ] = \sum_x \Pr[ X \!=\! x] \log_2(1 /\Pr[ X \!=\! x]). \label{eq:entropy}
  \end{align}
  The \emph{conditional entropy of $X$ conditioned on $Y$} is given by
  \begin{align}\label{eq:conditional_entropy}
    \HH[ X \mid Y ] &= \EE_Y[ \HH[ X \mid Y \!=\! y] ] \\
      &= \sum_{y}^{} \Pr[ Y \!=\! y] \HH[ X \mid Y \!=\! y].\notag
  \end{align}
\end{definition}

\begin{definition} \label{def:mutual0}
  Let $X$ and $Y$ be discrete random variables.
  The \emph{mutual information between $X$ and $Y$} is defined as
  \begin{align}
    \II[ X : Y ]
      &= \sum_{x,y}\Pr[x,y]\cdot \log\lb(\frac{\Pr[x,y]}{\Pr[x]\Pr[y]}\rb)
  \end{align}
\end{definition}

\begin{definition} \label{def:mutual}
  Let $X$, $Y$, and $Z$ be discrete random variables.
  The \emph{conditional mutual information of $X$ and $Y$} is defined as
  \begin{align}
    \II[ X : Y \mid Z ]
      &= \EE_Z[ \II[ X : Y \mid Z \!=\! z ]] \label{eq:mutual1} \\
      &= \HH[ X \mid Z ] - \HH[ X \mid Y, Z ] \label{eq:mutual2}.
  \end{align}
\end{definition}

\begin{lemma} \label{lem:entropy_conditioning}
  $\HH[ X \mid Y, Z ] \le \HH[ X \mid Y].$
\end{lemma}
\begin{lemma} \label{lem:mutual1}
  $\II[ X : Y \mid Z ] \le \HH[ X \mid Z ]$.
\end{lemma}

\begin{lemma}[Theorem~6.1 in \cite{ccbook}] \label{thm:shannon}
Consider any random variable $X$.
Every encoding of $X$ has expected length at least $\HH[X]$.
\end{lemma}

\begin{lemma}[Data Processing Inequality] \label{lem:dataprocessing}
  If random variables $X$, $Y$, and $Z$ form the Markov chain $X \ra Y \ra Z$, i.e., the conditional distribution of $Z$ depends only on $Y$ and is conditionally independent of $X$, then
  \[
    \II[ X : Y ] \ge \II[ X : Z ].
  \]
\end{lemma}

\begin{lemma}[Chain Rule of Mutual Information] \label{lem:chainrule}
Let $W$, $X$, $Y$, and $Z$ be jointly distributed random variables. Then
\begin{align}
 \II\lt[ X, Y : Z \ \md|\ W \rt] = \II\lt[ X : Z \ \md|\ W \rt]  + \II\lt[ Y : Z \ \md|\ W,X \rt].
\end{align}
\end{lemma}

\label{sec:tools}
\bibliographystyle{alpha}
\bibliographystyle{ACM-Reference-Format}
\bibliography{references}

\newcommand{\etalchar}[1]{$^{#1}$}
\begin{thebibliography}{BYJKS04}

\bibitem[AGG{\etalchar{+}}19]{DBLP:conf/spaa/AugustineGGHSKL19}
John Augustine, Mohsen Ghaffari, Robert Gmyr, Kristian Hinnenthal, Christian
  Scheideler, Fabian Kuhn, and Jason Li.
\newblock Distributed computation in node-capacitated networks.
\newblock In Christian Scheideler and Petra Berenbrink, editors, {\em The 31st
  {ACM} on Symposium on Parallelism in Algorithms and Architectures, {SPAA}
  2019, Phoenix, AZ, USA, June 22-24, 2019}, pages 69--79. {ACM}, 2019.

\bibitem[AGPV90]{DBLP:journals/jacm/AwerbuchGPV90}
Baruch Awerbuch, Oded Goldreich, David Peleg, and Ronen Vainish.
\newblock A trade-off between information and communication in broadcast
  protocols.
\newblock {\em J. {ACM}}, 37(2):238--256, 1990.

\bibitem[Awe85]{DBLP:journals/jacm/Awerbuch85}
Baruch Awerbuch.
\newblock Complexity of network synchronization.
\newblock {\em J. {ACM}}, 32(4):804--823, 1985.

\bibitem[BBCR13]{DBLP:journals/siamcomp/BarakBCR13}
Boaz Barak, Mark Braverman, Xi~Chen, and Anup Rao.
\newblock How to compress interactive communication.
\newblock {\em {SIAM} J. Comput.}, 42(3):1327--1363, 2013.

\bibitem[BEIK19]{DBLP:conf/wdag/BittonEIK19}
Shimon Bitton, Yuval Emek, Taisuke Izumi, and Shay Kutten.
\newblock Message reduction in the {LOCAL} model is a free lunch.
\newblock In {\em 33rd International Symposium on Distributed Computing, {DISC}
  2019, October 14-18, 2019, Budapest, Hungary}, pages 7:1--7:15, 2019.

\bibitem[BEO{\etalchar{+}}13]{DBLP:conf/focs/BravermanEOPV13}
Mark Braverman, Faith Ellen, Rotem Oshman, Toniann Pitassi, and Vinod
  Vaikuntanathan.
\newblock A tight bound for set disjointness in the message-passing model.
\newblock In {\em 54th Annual {IEEE} Symposium on Foundations of Computer
  Science, {FOCS} 2013, 26-29 October, 2013, Berkeley, CA, {USA}}, pages
  668--677. {IEEE} Computer Society, 2013.

\bibitem[BJKS04]{DBLP:journals/jcss/Bar-YossefJKS04}
Ziv Bar{-}Yossef, T.~S. Jayram, Ravi Kumar, and D.~Sivakumar.
\newblock An information statistics approach to data stream and communication
  complexity.
\newblock {\em J. Comput. Syst. Sci.}, 68(4):702--732, 2004.

\bibitem[BS07]{baswana}
Surender Baswana and Sandeep Sen.
\newblock A simple and linear time randomized algorithm for computing sparse
  spanners in weighted graphs.
\newblock {\em Random Struct. Algorithms}, 30(4):532--563, 2007.

\bibitem[BYJKS04]{bar2004information}
Ziv Bar-Yossef, Thathachar~S Jayram, Ravi Kumar, and D~Sivakumar.
\newblock An information statistics approach to data stream and communication
  complexity.
\newblock {\em Journal of Computer and System Sciences}, 68(4):702--732, 2004.

\bibitem[CHD18]{censor2018distributed}
Keren Censor-Hillel and Michal Dory.
\newblock Distributed spanner approximation.
\newblock In {\em Proceedings of the 2018 ACM Symposium on Principles of
  Distributed Computing}, pages 139--148, 2018.

\bibitem[CHKM12]{DBLP:conf/stoc/Censor-HillelHKM12}
Keren Censor{-}Hillel, Bernhard Haeupler, Jonathan~A. Kelner, and Petar
  Maymounkov.
\newblock Global computation in a poorly connected world: fast rumor spreading
  with no dependence on conductance.
\newblock In Howard~J. Karloff and Toniann Pitassi, editors, {\em Proceedings
  of the 44th Symposium on Theory of Computing Conference, {STOC} 2012, New
  York, NY, USA, May 19 - 22, 2012}, pages 961--970. {ACM}, 2012.

\bibitem[CSWY01]{chakrabarti2001informational}
Amit Chakrabarti, Yaoyun Shi, Anthony Wirth, and Andrew Yao.
\newblock Informational complexity and the direct sum problem for simultaneous
  message complexity.
\newblock In {\em Proceedings 42nd IEEE Symposium on Foundations of Computer
  Science}, pages 270--278. IEEE, 2001.

\bibitem[CT06]{info_book}
T.~Cover and J.A. Thomas.
\newblock {\em Elements of Information Theory, second edition}.
\newblock Wiley, 2006.

\bibitem[DGPV08]{derbel}
Bilel Derbel, Cyril Gavoille, David Peleg, and Laurent Viennot.
\newblock On the locality of distributed sparse spanner construction.
\newblock In {\em Proceedings of the twenty-seventh ACM symposium on Principles
  of distributed computing}, pages 273--282, 2008.

\bibitem[Elk07]{DBLP:conf/podc/Elkin07}
Michael Elkin.
\newblock A near-optimal distributed fully dynamic algorithm for maintaining
  sparse spanners.
\newblock In Indranil Gupta and Roger Wattenhofer, editors, {\em Proceedings of
  the Twenty-Sixth Annual {ACM} Symposium on Principles of Distributed
  Computing, {PODC} 2007, Portland, Oregon, USA, August 12-15, 2007}, pages
  185--194. {ACM}, 2007.

\bibitem[Elk17]{DBLP:conf/podc/Elkin17}
Michael Elkin.
\newblock A simple deterministic distributed {MST} algorithm, with near-optimal
  time and message complexities.
\newblock In Elad~Michael Schiller and Alexander~A. Schwarzmann, editors, {\em
  Proceedings of the {ACM} Symposium on Principles of Distributed Computing,
  {PODC} 2017, Washington, DC, USA, July 25-27, 2017}, pages 157--163. {ACM},
  2017.

\bibitem[EN19]{DBLP:journals/talg/ElkinN19}
Michael Elkin and Ofer Neiman.
\newblock Efficient algorithms for constructing very sparse spanners and
  emulators.
\newblock {\em {ACM} Trans. Algorithms}, 15(1):4:1--4:29, 2019.

\bibitem[Erd63]{erdos}
Paul Erd\"os.
\newblock Extremal problems in graph theory.
\newblock In {\em Proc. Symp. Theory of Graphs and its Applications}, page
  2936, 1963.

\bibitem[FGKO18]{DBLP:conf/spaa/FischerGKO18}
Orr Fischer, Tzlil Gonen, Fabian Kuhn, and Rotem Oshman.
\newblock Possibilities and impossibilities for distributed subgraph detection.
\newblock In Christian Scheideler and Jeremy~T. Fineman, editors, {\em
  Proceedings of the 30th on Symposium on Parallelism in Algorithms and
  Architectures, {SPAA} 2018, Vienna, Austria, July 16-18, 2018}, pages
  153--162. {ACM}, 2018.

\bibitem[FHW12]{DBLP:conf/soda/FrischknechtHW12}
Silvio Frischknecht, Stephan Holzer, and Roger Wattenhofer.
\newblock Networks cannot compute their diameter in sublinear time.
\newblock In Yuval Rabani, editor, {\em Proceedings of the Twenty-Third Annual
  {ACM-SIAM} Symposium on Discrete Algorithms, {SODA} 2012, Kyoto, Japan,
  January 17-19, 2012}, pages 1150--1162. {SIAM}, 2012.

\bibitem[FWY20]{DBLP:conf/innovations/FernandezW020}
Manuel Fernandez, David~P. Woodruff, and Taisuke Yasuda.
\newblock Graph spanners in the message-passing model.
\newblock In Thomas Vidick, editor, {\em 11th Innovations in Theoretical
  Computer Science Conference, {ITCS} 2020, January 12-14, 2020, Seattle,
  Washington, {USA}}, volume 151 of {\em LIPIcs}, pages 77:1--77:18. Schloss
  Dagstuhl - Leibniz-Zentrum f{\"{u}}r Informatik, 2020.

\bibitem[GK18]{DBLP:conf/wdag/GhaffariK18a}
Mohsen Ghaffari and Fabian Kuhn.
\newblock Distributed {MST} and broadcast with fewer messages, and faster
  gossiping.
\newblock In {\em 32nd International Symposium on Distributed Computing, {DISC}
  2018, New Orleans, LA, USA, October 15-19, 2018}, volume 121 of {\em LIPIcs},
  pages 30:1--30:12. Schloss Dagstuhl - Leibniz-Zentrum f{\"{u}}r Informatik,
  2018.

\bibitem[GP18]{DBLP:conf/wdag/GmyrP18}
Robert Gmyr and Gopal Pandurangan.
\newblock Time-message trade-offs in distributed algorithms.
\newblock In {\em 32nd International Symposium on Distributed Computing, {DISC}
  2018, New Orleans, LA, USA, October 15-19, 2018}, volume 121 of {\em LIPIcs},
  pages 32:1--32:18. Schloss Dagstuhl - Leibniz-Zentrum f{\"{u}}r Informatik,
  2018.

\bibitem[HHW18]{DBLP:conf/podc/HaeuplerHW18}
Bernhard Haeupler, D.~Ellis Hershkowitz, and David Wajc.
\newblock Round- and message-optimal distributed graph algorithms.
\newblock In {\em Proceedings of the 2018 {ACM} Symposium on Principles of
  Distributed Computing, {PODC} 2018, Egham, United Kingdom, July 23-27, 2018},
  pages 119--128, 2018.

\bibitem[HMS18]{DBLP:conf/podc/HaeuplerMS18}
Bernhard Haeupler, Jeet Mohapatra, and Hsin{-}Hao Su.
\newblock Optimal gossip algorithms for exact and approximate quantile
  computations.
\newblock In {\em Proceedings of the 2018 {ACM} Symposium on Principles of
  Distributed Computing, {PODC} 2018, Egham, United Kingdom, July 23-27, 2018},
  pages 179--188, 2018.

\bibitem[HPP{\etalchar{+}}15]{hegeman2015toward}
James~W Hegeman, Gopal Pandurangan, Sriram~V Pemmaraju, Vivek~B Sardeshmukh,
  and Michele Scquizzato.
\newblock Toward optimal bounds in the congested clique: Graph connectivity and
  mst.
\newblock In {\em Proceedings of the 2015 ACM Symposium on Principles of
  Distributed Computing}, pages 91--100, 2015.

\bibitem[IG17]{DBLP:conf/podc/IzumiG17}
Taisuke Izumi and Fran{\c{c}}ois~Le Gall.
\newblock Triangle finding and listing in {CONGEST} networks.
\newblock In Elad~Michael Schiller and Alexander~A. Schwarzmann, editors, {\em
  Proceedings of the {ACM} Symposium on Principles of Distributed Computing,
  {PODC} 2017, Washington, DC, USA, July 25-27, 2017}, pages 381--389. {ACM},
  2017.

\bibitem[KKT15]{DBLP:conf/podc/KingKT15}
Valerie King, Shay Kutten, and Mikkel Thorup.
\newblock Construction and impromptu repair of an {MST} in a distributed
  network with o(m) communication.
\newblock In Chryssis Georgiou and Paul~G. Spirakis, editors, {\em Proceedings
  of the 2015 {ACM} Symposium on Principles of Distributed Computing, {PODC}
  2015, Donostia-San Sebasti{\'{a}}n, Spain, July 21 - 23, 2015}, pages 71--80.
  {ACM}, 2015.

\bibitem[KNPR15]{soda15}
Hartmut Klauck, Danupon Nanongkai, Gopal Pandurangan, and Peter Robinson.
\newblock Distributed computation of large-scale graph problems.
\newblock In {\em Proceedings of the Twenty-Sixth Annual {ACM-SIAM} Symposium
  on Discrete Algorithms, {SODA} 2015, San Diego, CA, USA, January 4-6, 2015},
  pages 391--410, 2015.

\bibitem[KOS17]{DBLP:conf/wdag/KolOS17}
Gillat Kol, Rotem Oshman, and Dafna Sadeh.
\newblock Interactive compression for multi-party protocol.
\newblock In Andr{\'{e}}a~W. Richa, editor, {\em 31st International Symposium
  on Distributed Computing, {DISC} 2017, October 16-20, 2017, Vienna, Austria},
  volume~91 of {\em LIPIcs}, pages 31:1--31:15. Schloss Dagstuhl -
  Leibniz-Zentrum f{\"{u}}r Informatik, 2017.

\bibitem[KPP{\etalchar{+}}15]{jacm15}
Shay Kutten, Gopal Pandurangan, David Peleg, Peter Robinson, and Amitabh
  Trehan.
\newblock On the complexity of universal leader election.
\newblock {\em J. {ACM}}, 62(1):7:1--7:27, 2015.

\bibitem[LPP06]{DBLP:journals/dc/LotkerPP06}
Zvi Lotker, Boaz Patt{-}Shamir, and David Peleg.
\newblock Distributed {MST} for constant diameter graphs.
\newblock {\em Distributed Computing}, 18(6):453--460, 2006.

\bibitem[MK17]{DBLP:conf/icdcn/MashreghiK17}
Ali Mashreghi and Valerie King.
\newblock Time-communication trade-offs for minimum spanning tree construction.
\newblock In {\em Proceedings of the 18th International Conference on
  Distributed Computing and Networking, Hyderabad, India, January 5-7, 2017},
  page~8. {ACM}, 2017.

\bibitem[MK18]{DBLP:conf/wdag/MashreghiK18}
Ali Mashreghi and Valerie King.
\newblock Broadcast and minimum spanning tree with o(m) messages in the
  asynchronous {CONGEST} model.
\newblock In {\em 32nd International Symposium on Distributed Computing, {DISC}
  2018, New Orleans, LA, USA, October 15-19, 2018}, volume 121 of {\em LIPIcs},
  pages 37:1--37:17. Schloss Dagstuhl - Leibniz-Zentrum f{\"{u}}r Informatik,
  2018.

\bibitem[MK19]{DBLP:conf/wdag/MashreghiK19}
Ali Mashreghi and Valerie King.
\newblock Brief announcement: Faster asynchronous {MST} and low diameter tree
  construction with sublinear communication.
\newblock In Jukka Suomela, editor, {\em 33rd International Symposium on
  Distributed Computing, {DISC} 2019, October 14-18, 2019, Budapest, Hungary},
  volume 146 of {\em LIPIcs}, pages 49:1--49:3. Schloss Dagstuhl -
  Leibniz-Zentrum f{\"{u}}r Informatik, 2019.

\bibitem[MU05]{upfalmitzenmacher}
Michael Mitzenmacher and Eli Upfal.
\newblock {\em Probability and Computing: Randomized Algorithms and
  Probabilistic Analysis}.
\newblock Cambridge University Press, 2005.

\bibitem[Mun13]{munkres2013topology}
James Munkres.
\newblock {\em Topology: Pearson New International Edition}.
\newblock Pearson, 2013.

\bibitem[Pel00]{peleg_book}
David Peleg.
\newblock {\em Distributed Computing: A Locality-Sensitive Approach}.
\newblock Society for Industrial and Applied Mathematics, 2000.

\bibitem[PPP{\etalchar{+}}17]{DBLP:conf/wdag/PaiPPR017}
Shreyas Pai, Gopal Pandurangan, Sriram~V. Pemmaraju, Talal Riaz, and Peter
  Robinson.
\newblock Symmetry breaking in the congest model: Time- and message-efficient
  algorithms for ruling sets.
\newblock In Andr{\'{e}}a~W. Richa, editor, {\em 31st International Symposium
  on Distributed Computing, {DISC} 2017, October 16-20, 2017, Vienna, Austria},
  volume~91 of {\em LIPIcs}, pages 38:1--38:16. Schloss Dagstuhl -
  Leibniz-Zentrum f{\"{u}}r Informatik, 2017.

\bibitem[PPPR21]{DBLP:conf/podc/PaiPP021}
Shreyas Pai, Gopal Pandurangan, Sriram~V. Pemmaraju, and Peter Robinson.
\newblock Can we break symmetry with o(m) communication?
\newblock In Avery Miller, Keren Censor{-}Hillel, and Janne~H. Korhonen,
  editors, {\em {PODC} '21: {ACM} Symposium on Principles of Distributed
  Computing, Virtual Event, Italy, July 26-30, 2021}, pages 247--257. {ACM},
  2021.

\bibitem[PPS16]{DBLP:conf/sirocco/PanduranganPS16}
Gopal Pandurangan, David Peleg, and Michele Scquizzato.
\newblock Message lower bounds via efficient network synchronization.
\newblock In Jukka Suomela, editor, {\em Structural Information and
  Communication Complexity - 23rd International Colloquium, {SIROCCO} 2016,
  Helsinki, Finland, July 19-21, 2016, Revised Selected Papers}, volume 9988 of
  {\em Lecture Notes in Computer Science}, pages 75--91, 2016.

\bibitem[PRS17]{DBLP:conf/stoc/Pandurangan0S17}
Gopal Pandurangan, Peter Robinson, and Michele Scquizzato.
\newblock A time- and message-optimal distributed algorithm for minimum
  spanning trees.
\newblock In {\em Proceedings of the 49th Annual {ACM} {SIGACT} Symposium on
  Theory of Computing, {STOC} 2017, Montreal, QC, Canada, June 19-23, 2017},
  pages 743--756, 2017.

\bibitem[PRS18]{DBLP:conf/spaa/Pandurangan0S18}
Gopal Pandurangan, Peter Robinson, and Michele Scquizzato.
\newblock On the distributed complexity of large-scale graph computations.
\newblock In Christian Scheideler and Jeremy~T. Fineman, editors, {\em
  Proceedings of the 30th on Symposium on Parallelism in Algorithms and
  Architectures, {SPAA} 2018, Vienna, Austria, July 16-18, 2018}, pages
  405--414. {ACM}, 2018.

\bibitem[PRVY19]{DBLP:conf/innovations/ParterRVY19}
Merav Parter, Ronitt Rubinfeld, Ali Vakilian, and Anak Yodpinyanee.
\newblock Local computation algorithms for spanners.
\newblock In Avrim Blum, editor, {\em 10th Innovations in Theoretical Computer
  Science Conference, {ITCS} 2019, January 10-12, 2019, San Diego, California,
  {USA}}, volume 124 of {\em LIPIcs}, pages 58:1--58:21. Schloss Dagstuhl -
  Leibniz-Zentrum f{\"{u}}r Informatik, 2019.

\bibitem[PS89]{DBLP:journals/jgt/PelegS89}
David Peleg and Alejandro~A. Sch{\"{a}}ffer.
\newblock Graph spanners.
\newblock {\em Journal of Graph Theory}, 13(1):99--116, 1989.

\bibitem[PVZ16]{DBLP:journals/siamcomp/PhillipsVZ16}
Jeff~M. Phillips, Elad Verbin, and Qin Zhang.
\newblock Lower bounds for number-in-hand multiparty communication complexity,
  made easy.
\newblock {\em {SIAM} J. Comput.}, 45(1):174--196, 2016.

\bibitem[PY19a]{DBLP:conf/soda/ParterY19a}
Merav Parter and Eylon Yogev.
\newblock Distributed algorithms made secure: {A} graph theoretic approach.
\newblock In Timothy~M. Chan, editor, {\em Proceedings of the Thirtieth Annual
  {ACM-SIAM} Symposium on Discrete Algorithms, {SODA} 2019, San Diego,
  California, USA, January 6-9, 2019}, pages 1693--1710. {SIAM}, 2019.

\bibitem[PY19b]{DBLP:conf/podc/ParterY19}
Merav Parter and Eylon Yogev.
\newblock Secure distributed computing made (nearly) optimal.
\newblock In Peter Robinson and Faith Ellen, editors, {\em Proceedings of the
  2019 {ACM} Symposium on Principles of Distributed Computing, {PODC} 2019,
  Toronto, ON, Canada, July 29 - August 2, 2019}, pages 107--116. {ACM}, 2019.

\bibitem[RY20]{ccbook}
Anup Rao and Amir Yehudayoff.
\newblock {\em Communication Complexity: and Applications}.
\newblock Cambridge University Press, 2020.

\bibitem[SHK{\etalchar{+}}12]{DBLP:journals/siamcomp/SarmaHKKNPPW12}
Atish~Das Sarma, Stephan Holzer, Liah Kor, Amos Korman, Danupon Nanongkai,
  Gopal Pandurangan, David Peleg, and Roger Wattenhofer.
\newblock Distributed verification and hardness of distributed approximation.
\newblock {\em {SIAM} J. Comput.}, 41(5):1235--1265, 2012.

\bibitem[Suo13]{jukkaLocalAlgorithms}
Jukka Suomela.
\newblock Survey of local algorithms.
\newblock {\em {ACM} Comput. Surv.}, 45(2):24:1--24:40, 2013.

\bibitem[WZ17]{DBLP:journals/dc/WoodruffZ17}
David~P. Woodruff and Qin Zhang.
\newblock When distributed computation is communication expensive.
\newblock {\em Distributed Computing}, 30(5):309--323, 2017.

\end{thebibliography}

\end{document}